\documentclass[journal]{IEEEtran}
\usepackage{amsmath,amssymb,physics,tabularx,graphicx,multirow,hhline,algorithm,mathtools,algpseudocode,varwidth,color,soul,amsthm,cite,bbm}
\usepackage[skins,breakable,hooks,theorems]{tcolorbox}

\algnewcommand{\algorithmicforeach}{\textbf{for each}}
\algdef{SE}[FOR]{ForEach}{EndForEach}[1]
  {\algorithmicforeach\ #1\ \algorithmicdo}
  {\algorithmicend\ \algorithmicforeach}
\newsavebox{\ieeealgbox}

\ifCLASSINFOpdf
\else
\fi

\newtheorem{definition}{Definition}

\tcbset{
  highlight math style={
    colback=yellow,
    arc=0pt,
    outer arc=0pt,
    boxrule=0pt,
    top=2pt,
    bottom=2pt,
    left=2pt,
    right=2pt,
  }
}

\ifCLASSOPTIONcompsoc \usepackage[caption=false,font=normalsize,labelfont=sf,textfont=sf]{subfig}
\else
\usepackage[caption=false,font=footnotesize]{subfig}
\fi

\begin{document}
\algnewcommand{\algorithmicgoto}{\textbf{go to}}%
\algnewcommand{\Goto}[1]{\algorithmicgoto~\ref{#1}}%
\title{Transient Stability Assessment of Networked Microgrids Using Neural Lyapunov Methods}

\author{Tong Huang,~\IEEEmembership{Student Member,~IEEE,}
		Sicun Gao,
        and Le Xie,~\IEEEmembership{Senior Member,~IEEE}



        }



\maketitle

\begin{abstract}
This paper proposes a novel transient stability assessment tool for networked microgrids based on neural Lyapunov methods. Assessing transient stability is formulated as a problem of estimating the dynamic security region of networked microgrids. We leverage neural networks to learn a local Lyapunov function in the state space. The largest security region is estimated based on the learned neural Lyapunov function, and it is used for characterizing disturbances that the networked microgrids can tolerate. The proposed method is tested and validated in a grid-connected microgrid, three networked microgrids with mixed interface dynamics, and the IEEE 123-node feeder. Case studies suggest that the proposed method can address networked microgrids with heterogeneous interface dynamics, and in comparison with conventional methods that are based on quadratic Lyapunov functions, can characterize the security regions with much less conservativeness. 
\end{abstract}


\IEEEpeerreviewmaketitle{}

\section{Introduction} 
\label{sec:introduction}
\begin{figure}[b]
    \centering
    \includegraphics[width = 3in]{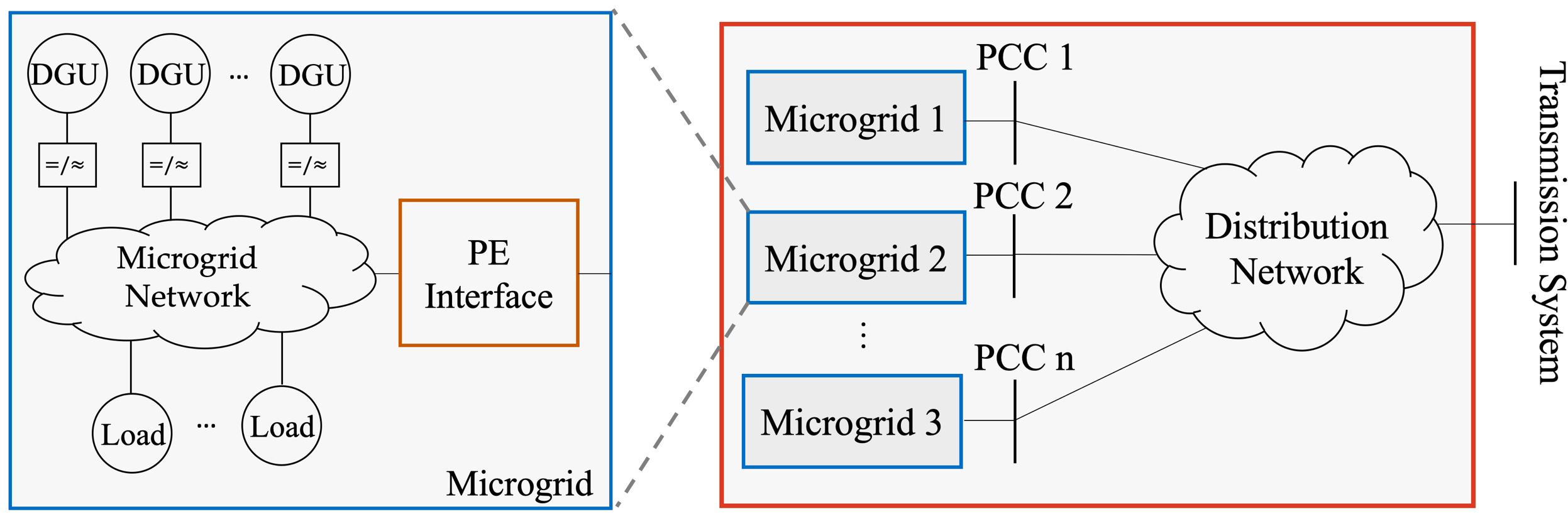}
    \caption{A microgrid-based distribution system: inside the left blue box shows the physical structure of a microgrid.}
    \label{fig:networked_MG}
\end{figure}
The past decade has witnessed increasing deployment of distributed energy resources (DERs) in the electric distribution grid. DERs play a crucial role of decarbonizing the energy sector and enhancing the resilience of the grid. However, deepening penetration of DERs leads to unprecedented complexity for distribution system operation in monitoring, control, and protection. One promising solution to managing massive integration of DERs is to reconfigure the distribution system as \emph{networked microgrids} shown in Figure \ref{fig:networked_MG}. A microgrid packages interconnected Distributed Generation Units (DGUs) and loads which are regulated locally by the Microgrid Central Controller (MGCC) \cite{ZhangLarge}. The microgrid has a power-electronic (PE) interface \cite{ZhangLarge,7322287} that physically connects to its host distribution system via a point of common coupling (PCC). Microgrids are networked with each other through PCCs and distribution lines. With such a configuration, instead of managing massive DGUs at grid edges, a Distribution System Operator (DSO) only needs to coordinate a few PE interfaces of microgrids \cite{ZhangLarge}, by which the system management complexity at the DSO level is significantly reduced. Reference \cite{7947305} reports a real-world demonstration of networked microgrids.

Given the microgrid-based distribution system, an essential function of its Distribution Management System (DMS) is to assess security of networked microgrids. Specifically, such a function is expected to include the static security assessment (SSA) and transient stability assessment (TSA). The SSA scrutinizes if physical variables of networked microgrids in the quasi-steady-state time scale are within normal ranges. It is typically considered as optimization constraints when researchers develop coordination strategies of networked microgrids \cite{8536454} for grid resilience enhancement and economical efficiency maximization. The TSA examines the dynamic behaviors of networked microgrids in a faster time scale. The TSA tool aims to characterize (large) disturbances that the networked microgrids can tolerate. Such characterization allows for efficient design and planning of the microgrid-based distribution systems \cite{HICSS2021Huang}, and it also enables a DSO to maintain situational awareness in real-time operation \cite{HICSS2021Huang}. This paper focuses on assessing transient stability of \emph{networked} microgrids. Such a topic concerns the DSO, because excessive energy transactions among microgrids may lead to stability issues, even though each individual microgrid is stabilized by its local MGCC \cite{ZhangLarge}.

There are several approaches to the design of TSA tools for networked microgrids. One may tailor the TSA development in bulk transmission systems for microgrid application. A prevailing TSA method in transmission systems is based on time-domain simulation \cite{1338120}. In such a method, the system responses are simulated given all credible contingencies \cite{1338120}. System security is evaluated by examining the responses obtained. This method can be tailored to screen out critical contingencies in networked microgrids. However, it cannot certify stability rigorously, as by definition, stability \cite{slotine1991applied} requires one to examine system responses under infinite number of disturbances, which is impossible for time-domain simulation.
Another TSA method developed for the transmission grid is the energy function method \cite{41298,481632,MIT}. By assuming transmission lines are lossless, this method aims to construct an energy function that can certify stability. While the lossless-line assumption is plausible in transmission systems, it does not hold in networked microgrids due to large R/X ratios of distribution lines \cite{ZhangLarge}, thereby resulting in non-existence of the energy function in networked microgrids \cite{41298}. Reference \cite{41298} constructs a quadratic Lyapunov function which can be used for TSA of a power system with line loss. However, the DSO tool developed based on \cite{41298} may be overly conservative. Besides the TSA tools developed for transmission systems, References \cite{7322287,ZhangLarge} develop stability assessment tools specifically for networked microgrids. Reference \cite{7322287} has proposed a framework capable of assessing the small-signal stability of networked microgrids in a distributed manner, but it cannot certify the stability when large disturbances occur. Reference \cite{ZhangLarge} utilizes linear matrix inequalities (LMIs) in order to certify global asymptotic stability of networked microgrids. The framework proposed in \cite{ZhangLarge} requires a special form of interface dynamics and it cannot characterize disturbances that can be tolerated by networked microgrids when global asymptotic stability is not guaranteed. 

In this paper, we develop a TSA tool for networked microgrids using machine learning-based Neural Lyapunov Methods. Assessing transient stability is formulated as a problem of estimating the security region of networked microgrids. We leverage neural networks to learn a local Lyapunov function in the state space. The optimal security region is estimated based on the Lyapunov function learned, and is used for characterizing disturbances that the networked microgrids can tolerate. The proposed TSA tool has the following merits: 1) It can provide less conservative characterization of disturbances that can be tolerated by networked microgrids, compared with methods based on quadratic Lyapunov functions; and 2) It can assess the transient stability of networked microgrids with heterogeneous interface dynamics.

The rest of this paper is organized as follows: Section \ref{sec:dynamics_of_power_electronic_interfaced_microgrids} describes the dynamics of networked microgrids; Section \ref{sec:neural_lyapunov_method_to_transient_stability_assessment} presents the neural Lyapunov method to TSA; and Section \ref{sec:numerical_experiments} tests and validates the tool in three numerical experiments; and Section \ref{sec:conclusion} concludes the paper and points out future direction.
\section{Dynamics of Microgrids with PE Interfaces} 
\label{sec:dynamics_of_power_electronic_interfaced_microgrids}

With the physical configuration of the networked microgrids in Figure \ref{fig:networked_MG}, the dynamics that a microgrid exhibits at the DSO-level control are mainly determined by the control strategy deployed at its power electronics interface \cite{7322287,ZhangLarge}. One promising control strategy for the PE interface is the droop method \cite{ZhangLarge}. 
Such a method does not need explicit communication between microgrid interfaces in order to achieve load sharing, allowing for distributed implementation \cite{ZhangLarge,7862299,8892668,8390902}.
For the droop method deployed at a microgrid interface, some local signals are leveraged as the power balance indicators \cite{ZhangLarge}, and the interface response is tuned according to the measurements of these signals. Common selections of these local signals include frequency, voltage magnitude and voltage angle that are measured at the microgrid PCC. Specifically, the \emph{frequency droop control} takes the frequency as the balance indicator for real power \cite{8892668}, while the \emph{angle droop control} considers the voltage phase angle as the balance indicator \cite{ZhangLarge,7862299,8390902}. 

For the $n$ networked microgrids in Figure \ref{fig:networked_MG}, without loss of generality, suppose that the angle droop control is deployed in the $k$-th microgrid's PE interface, where $k = 1, 2, \ldots, n$. The interface dynamics of the $k$-th microgrid are \cite{ZhangLarge,7322287}
\begin{subequations}\label{eq:diff_eq}
    \begin{align}
        &M_{\text{a}k}\dot{\delta}_k' + \delta'_k = D_{\text{a}k}(P_k^* - P_k)\\
        &M_{\text{v}k}\dot{E}_k' + E_k' = D_{\text{v}k}(Q_k^* - Q_k)
    \end{align} 
\end{subequations}
where $\delta_k'$ and $E_k'$ are \emph{deviations} of voltage phase angle $\delta_k$ and voltage magnitude $E_k$ from their steady state values $\delta_k^*$ and $E_k^*$ at the $k$-th PCC, respectively, i.e., $\delta'_k:=\delta_k - \delta_k^*$ and $E'_k:= E_k - E_k^*$; $M_{\text{a}k}$ and $M_{\text{v}k}$ are tracking time constants; $D_{\text{a}k}$ and $D_{\text{v}k}$ are droop gains; $P_k$ and $Q_k$ denote real and reactive power injections to PCC $k$; and $P_k^*$ and $Q_k^*$ are the steady-state injections of real and reactive power at the $k$-th PCC \cite{ZhangLarge}. The $n$ microgrids are networked via distribution network which introduces constrains
\begin{subequations}\label{eq:power_flow}
    \begin{align}
        &P_k - G_{kk}E_k^2 - \sum_{i\ne k}E_kE_iY_{ki}\cos{(\delta_{ki}-\sigma_{ki})} = 0\\
        &Q_k + B_{kk}E_k^2 - \sum_{i\ne k}E_kE_iY_{ki}\sin{(\delta_{ki}-\sigma_{ki})} =0, \forall k,
    \end{align} 
\end{subequations}
where $\delta_{ki}=\delta_k - \delta_i$; $G_{kk}+\mathbbm{j}B_{kk}$ is the $k$-th diagonal entry in the admittance matrix of the distribution network; and $Y_{ki}\angle \sigma_{ki}$ is the $(k,i)$-th entry of the admittance matrix. The steady-state values $\delta_k^*$, $E_k^*$, $P_k^*$ and $Q_k^*$ are designed based on economic dispatch and they satisfy the following equality constrains:
\begin{subequations}
    \begin{align}
        &P_k^* - G_{kk}E_k^{*2} - \sum_{i\ne k}E_k^*E_i^*Y_{ki}\cos{(\delta_{ki}^*-\sigma_{ki})} = 0\\
        &Q_k^* + B_{kk}E_k^{*2} - \sum_{i\ne k}E_k^*E_i^*Y_{ki}\sin{(\delta_{ki}^*-\sigma_{ki})} =0, \forall k,
    \end{align} 
\end{subequations}
where $\delta_{ki}^* = \delta_k^* - \delta_i^*$.
Differential equations \eqref{eq:diff_eq} with algebraic equations \eqref{eq:power_flow} characterize the dynamics of the $n$ networked microgrids, and their compact form is
\begin{equation} \label{eq:f_x}
    \dot{\mathbf{x}} = \mathbf{f}(\mathbf{x})
\end{equation}
where $\mathbf{x}=[\delta'_1, \delta'_2, \ldots, \delta'_n, E_1', E_2', \ldots, E_n']$; and $\mathbf{f}(\cdot)$ is determined by \eqref{eq:diff_eq} and \eqref{eq:power_flow}. Note that the equilibrium point $\mathbf{o}$ of the dynamic system \eqref{eq:f_x} is the origin of the state space.

If $M_{\text{v}k}\gg M_{\text{a}k}$, the \emph{time-scale separation} can be assumed \cite{ZhangLarge,HICSS2021Huang}. In such a case, the voltage deviation $E'_k$ evolves much slower than the phase angle deviation $\delta_k'$ and, therefore, $E'_k$ is assumed to be constant \cite{ZhangLarge}. Furthermore, if only angular stability is of interest, the dynamics of the $n$ networked microgrids can be described by
\begin{equation}\label{eq:time_scale}
        M_{\text{a}k}\dot{\delta}_k' + \delta'_k = D_{\text{a}k}(P_k^* - P_k), \forall k,
\end{equation}
where $P_k = G_{kk}E_k^{*2} + \sum_{i\ne k}E_k^*E_i^*Y_{ki}\cos{(\delta'_{ki}+\delta_{ki}^*-\sigma_{ki})}$.
The compact form of \eqref{eq:time_scale} can be also expressed as \eqref{eq:f_x} where $\mathbf{x}$ and $\mathbf{f}(\cdot)$ should be revised accordingly. Besides, with the time-scale separation assumption, if the frequency droop control is deployed in the $j$-th microgrid, the $j$-th differential equation in \eqref{eq:time_scale} is replaced by
\begin{equation*}
    \dot{\delta}_j' = \omega_j', \quad M_{\text{f}j}\dot{\omega}_j' + D_{\text{f}j} \omega_j' = P_j^* - P_j
\end{equation*}
where $\omega'_j$ denotes the frequency deviation from its nominal value at the $j$-th PCC; $M_{\text{f}j}$ and $D_{\text{f}j}$ are the emulated inertia and damping coefficients, respectively; and $P_j = G_{jj}E_j^{*2} + \sum_{i\ne j}E_j^*E_i^*Y_{ji}\cos{(\delta'_{ji}+\delta_{ji}^*-\sigma_{ji})}$.

With the networked microgrids \eqref{eq:f_x} and its equilibrium point $\mathbf{o}$, a DSO may have the following two questions \cite{HICSS2021Huang}: 1) Is $\mathbf{o}$ asymptotically stable? 2) How ``large'' are the disturbances that the networked microgrids can tolerate?
The transient stability assessment framework proposed in this paper aims to answer these two questions.



\section{Neural Lyapunov Methods} 
\label{sec:neural_lyapunov_method_to_transient_stability_assessment}
This section answers the two DSO's questions. We first point out the asymptotic stability of networked microgrids can be certified by the Lyapunov linearization method \cite{slotine1991applied} and formulate the second DSO's question as the one of estimating a security region of networked microgrids. Then an optimal security region is estimated via learning a Lyapunov function. Finally, how to empirically tune the parameters of proposed algorithms is discussed. 
\subsection{Asymptotic Stability Check and Security Region} 
\label{sub:asymptotic_stability_check_and_security_region_estimation}
Given the networked microgrids \eqref{eq:f_x} and its equilibrium $\mathbf{o}$, the Lyapunov linearization method \cite{slotine1991applied} suggests the asymptotic stability of $\mathbf{o}$ can be determined by examining the linearized version of \eqref{eq:f_x}, i.e.,
\begin{equation}\label{eq:Ax}
    \dot{\mathbf{x}} = A\mathbf{x}.
\end{equation}
In \eqref{eq:Ax}, $A\in\mathbb{R}^{m\times m}$ is a system matrix, where $m$ is the length of the state vector $\mathbf{x}$. The system matrix $A$ is obtained by linearizing \eqref{eq:f_x} around its equilibrium point $\mathbf{o}$ based on the linearization technique. Suppose that matrix $A$ has $m$ eigenvalues $\lambda_1, \lambda_2, \ldots, \lambda_m$. The equilibrium point $\mathbf{o}$ of \eqref{eq:f_x} is asymptotically stable \cite{slotine1991applied}, if
\begin{equation} \label{eq:stability_check}
    \Re({\lambda_i})<0 \quad \forall i=1, 2,\ldots m.
\end{equation} 
Condition \eqref{eq:stability_check} answers the first question raised in Section \ref{sec:dynamics_of_power_electronic_interfaced_microgrids}.

For the second DSO's question, a \emph{security region} can be leveraged to characterize the disturbances that the networked microgrids \eqref{eq:f_x} operating at $\mathbf{o}$ are able to tolerate. The definition of a security region is as follows \cite{HICSS2021Huang}:
\begin{definition}\label{def:SecurityRegion}
    $\mathcal{S}\subseteq \mathbb{R}^{m}$ is a security region if
    \begin{equation*}
        \mathbf{x}(0)\in \mathcal{S} \Longrightarrow \mathbf{x}(\infty)=\mathbf{0}_{m} \land \forall t(t>0\implies \mathbf{x}(t)\in \mathcal{S}).
    \end{equation*}
\end{definition}
In Definition \ref{def:SecurityRegion}, $\mathbf{x}(0)$ is resulting from the microgrid interconnection-level events, say, topology changes of distribution system network, and one of the microgrids enters an islanded/grid-connected mode; and $\mathbf{0}_{m}$ denotes the origin of the state space with $m$ states. Definition \ref{def:SecurityRegion} essentially says that the system trajectory starting in the security region $\mathcal{S}$ will stay in $\mathcal{S}$ and tends to the equilibrium point $\mathbf{o}$. The second DSO's question can be answered if such an security region is obtained.

A security region $\mathcal{S}$ can be estimated based on a system behavior-summary function, i.e., a Lyapunov function, in conjunction with the Local Invariant Set Theorem \cite{slotine1991applied}. The Lyapunov function is given by the following definition \cite{HICSS2021Huang}:
\begin{definition} \label{def:Lyapunov_function}
    A continuous differentiable scalar function $V(\mathbf{x})$ is a Lyapunov function, if, in a region $\mathcal{B}_u:=\{\mathbf{x}\in\mathbb{R}^m| u>0, \norm{\mathbf{x}}_2^2<u^2\}$, 1) $V$ is positive definite in $\mathcal{B}_u$, and 2) $\dot{V}$ is negative definite in $\mathcal{B}_u$.
\end{definition}

Once a legitimate Lyapunov function $V(\mathbf{x})$ is available, a region $\mathcal{S}_d$ can be found by
\begin{equation} \label{eq:S_d}
    \mathcal{S}_d = \{\mathbf{x}\in\mathcal{B}_u|d>0,V(\mathbf{x})<d\}.
\end{equation}
The region $\mathcal{S}_d$ is an invariant set due to the decreasing nature of the Lyapunov function $V(\mathbf{x})$. Besides, the Invariant Set Theorem \cite{slotine1991applied} suggests that with the Lyapunov function $V(\mathbf{x})$, a system trajectory $\mathbf{x}(t)$ starting in $\mathcal{S}_d$ converges to the origin of the state space. Therefore, the region $\mathcal{S}_d$ is a security region. In order to characterize the disturbances that the networked microgrids can tolerate, the remaining questions are: 1) How to find a legitimate Lyapunov function in a valid region $\mathcal{B}_u$; and 2) with a Lyapunov function valid in $\mathcal{B}_u$, how to make the security region $\mathcal{S}_d$ as large as possible by tuning $d$ in \eqref{eq:S_d}. These two questions are addressed in Sections \ref{sec:neural_lyapunov_method_to_transient_stability_assessment}-B and \ref{sec:neural_lyapunov_method_to_transient_stability_assessment}-C.

\subsection{Learning Lyapunov Function from State Space} 
\label{sub:learning_lyapunov_function}

\subsubsection{Lyapunov Function with Neural-network Structure} \label{sub:lyapunov_function_with_neural_network_structure}
We assume that a Lyapunov function candidate is a neural network. The neural network has a hidden layer and an output layer. 
The input of the hidden layer is the state vector $\mathbf{x}\in \mathbb{R}^m$ and the output is a vector $\mathbf{v}_1\in\mathbb{R}^p$ where $p$ is the number of neurons in the hidden layer. Function $g_1:\mathbb{R}^m \rightarrow \mathbb{R}^p$ describes the relationship between $\mathbf{x}$ and $\mathbf{v}_1$ and its definition is
\begin{equation} \label{eq:hidden_layer}
    \mathbf{v}_1 =g_1(\mathbf{x}):= \texttt{tanh}(W_1\mathbf{x}+\mathbf{b}_1)
\end{equation}
where $W_1\in\mathbb{R}^{p\times m}$; $\mathbf{b}_1 \in \mathbb{R}^p$; and $\texttt{tanh}(\cdot)$ is an entry-wised hyperbolic function \cite{HICSS2021Huang}. Furthermore, we define an intermediate vector $\mathbf{c}_1 = [c_{1,1}, c_{1, 2}, \ldots, c_{1,p}]^{\top}$ for the hidden layer by $\mathbf{c}_1=W_1\mathbf{x}+\mathbf{b}_1$.
For the output layer, its input is vector $\mathbf{v}_1$ and its output is $V_{\boldsymbol{\theta}}\in\mathbb{R}$ which is interpreted as the Lyapunov candidate evaluated at vector $\mathbf{x}$. $V_{\boldsymbol{\theta}}$ is related with $\mathbf{v}_1$ via function $g_2:\mathbb{R}^p \rightarrow \mathbb{R}$ defined by
\begin{equation} \label{eq:output_layer}
    V_{\boldsymbol{\theta}} = g_2(\mathbf{v}_1):=\tanh(W_2 \mathbf{v}_1 +b_2)
\end{equation}
where $W_2\in\mathbb{R}^{1\times p}$; and $b_2 \in \mathbb{R}$. The intermediate variable $c_2$ associated with the output layer is defined by $c_2 = W_2 \mathbf{v}_1 +b_2$.
In sum, the Lyapunov function candidate is
\begin{equation} \label{eq:summary}
     V_{\boldsymbol{\theta}}(\mathbf{x})=g_2(g_1(\mathbf{x})).
 \end{equation} Denote by $\boldsymbol{\theta}$ the vector that consists of all unknown entries in $W_1$, $\mathbf{b}_1$, $W_2$, and $b_2$. The subscript of $V_{\boldsymbol{\theta}}$ indicates that the Lyapunov function candidate depends on $\boldsymbol{\theta}$.

\subsubsection{Lyapunov Risk Minimization} 
\label{sub:LyapunovRiskMinimization}
We proceed to tune $\boldsymbol{\theta}$ such that $V_{\boldsymbol{\theta}}(\mathbf{x})$ in \eqref{eq:summary} meets the two conditions in Definition \ref{def:Lyapunov_function}. Suppose that there are $q$ state vectors $\mathbf{x}_1, \mathbf{x}_2, \ldots, \mathbf{x}_q$. Let set $\mathcal{X}$ collect these $q$ vector samples. To tune $\boldsymbol{\theta}$, we introduce a cost function called (empirical) Lyapunov risk, i.e.,
\begin{equation} \label{eq:Lyapunov_risk}
    \begin{aligned}
        R_q(\boldsymbol{\theta}) &= \frac{\alpha}{q}\sum_{i=1}^q\left(\texttt{ReLU}(-V_{\boldsymbol{\theta}}(\mathbf{x}_i))\right)\\
        &+\frac{\beta}{q}\sum_{i=1}^q\left(\texttt{ReLU}(\dot{V}_{\boldsymbol{\theta}}(\mathbf{x}_i)+\tau)\right)
        + \gamma V^2_{\boldsymbol{\theta}}(\mathbf{0}_m)
    \end{aligned}
\end{equation}
where the tunable parameters $\alpha$, $\beta$, $\gamma$ and $\tau$ are positive scalars; $\texttt{ReLU}(\cdot)$ denotes the rectified linear unit; and $\dot{V}_{\boldsymbol{\theta}}$ is given by \cite{HICSS2021Huang}
\begin{equation} \label{eq:time_derivative}
    \dot{V}_{\boldsymbol{\theta}} = \frac{\partial V_{\boldsymbol{\theta}}}{\partial \mathbf{x}}\mathbf{f}(\mathbf{x})=\frac{\partial V_{\boldsymbol{\theta}}}{\partial c_2}\frac{\partial c_2}{\partial \mathbf{v}_1}\frac{\partial \mathbf{v}_1}{\partial \mathbf{c}_1}\frac{\partial \mathbf{c}_1}{\partial \mathbf{x}}\mathbf{f}(\mathbf{x}).
\end{equation}
In \eqref{eq:time_derivative}, the dynamics $\mathbf{f}(\mathbf{x})$ is provided in \eqref{eq:f_x};
\begin{subequations}
    \begin{align*}
        &\frac{\partial V_{\boldsymbol{\theta}}}{\partial c_2} = 1 - V_{\boldsymbol{\theta}}^2;\frac{\partial c_2}{\partial \mathbf{v}_1} = W_2; \frac{\partial \mathbf{c}_1}{\partial \mathbf{x}} = W_1; \text{and}\\
        &\frac{\partial \mathbf{v}_1}{\partial \mathbf{c}_1}=\text{diag}\left(1-\tanh^2(c_{1,1}),\ldots, 1-\tanh^2(c_{1,p}) \right).
    \end{align*}
\end{subequations}

The interpretation of the Lyapunov risk \eqref{eq:Lyapunov_risk} is presented as follows. In \eqref{eq:Lyapunov_risk}, The first ``\texttt{ReLU}'' term incurs positive penalty if $V_{\boldsymbol{\theta}}(\mathbf{x}_i)$ is negative. The second ``\texttt{ReLU}'' term results to positive penalty if $\dot{V}_{\boldsymbol{\theta}}(\mathbf{x}_i)$ is greater than $-\tau$. If the evaluation of $V_{\boldsymbol{\theta}}$ at the origin of the state space is not zero, the Lyapunov risk also increases according to \eqref{eq:Lyapunov_risk}. Parameters $\alpha$, $\beta$, $\gamma$ and $\tau$ determine the importance of the three terms of \eqref{eq:Lyapunov_risk} and their tuning procedure is discussed in Section \ref{sub:parameter_tunning}.

Given the training set $\mathcal{X}$, in order to find a Lyapunov function valid in $\mathcal{B}_u$, unknown parameters $\boldsymbol{\theta}$ should be chosen such that the Lyapunov risk $R_q(\boldsymbol{\theta})$ is minimized, viz.
\begin{equation} \label{eq:optimization}
    \min_{\boldsymbol{\theta}}R_q(\boldsymbol{\theta}).
\end{equation}
The gradient decent algorithm can be leveraged to solve \eqref{eq:optimization}. Algorithm \ref{alg:learning} presents a procedure to update $\boldsymbol{\theta}$, where $\boldsymbol{\theta}_0$ is the initial guess of $\boldsymbol{\theta}$; $r\in \mathbb{Z}_+$ denotes the times of updating $\boldsymbol{\theta}$; and the positive scalar $\eta$ is the learning rate. Note that merely using Algorithm \ref{alg:learning} to update $\boldsymbol{\theta}$ is not sufficient even with a large $r$. One reason is that $\mathcal{X}$ solely covers a finite number of training samples in $\mathcal{B}_u$. With the $\boldsymbol{\theta}$ obtained by Algorithm \ref{alg:learning} based on $\mathcal{X}$, it is possible that one or both of the two conditions in Definition \ref{def:Lyapunov_function} are violated in some part of $\mathcal{B}_u$ that is not included in $\mathcal{X}$. This issue is addressed in Section \ref{sub:falsification}.

\begin{algorithm}
    \caption{Lyapunov Risk Minimization} \label{alg:learning}
        \begin{algorithmic}[1]
            \Function{\tt{MinRisk}}{$\boldsymbol{\theta}_0, \mathcal{X}, \mathbf{f}, p, r, \eta, \alpha, \beta, \gamma, \tau$}
            \State $\boldsymbol{\theta}\leftarrow \boldsymbol{\theta}_0$
            \While{$i \le r$}
            \State Update $V_{\boldsymbol{\theta}}$ and $\dot{V}_{\boldsymbol{\theta}}$ by \eqref{eq:summary}, \eqref{eq:time_derivative} with $\boldsymbol{\theta}$
            \State Compute $R_{\abs{\mathcal{X}}, \rho}(\boldsymbol{\theta})$ via \eqref{eq:Lyapunov_risk} over $\mathcal{X}$
            \State $\boldsymbol{\theta}\leftarrow \boldsymbol{\theta}-\eta \nabla_{\boldsymbol{\theta}}R_{\abs{\mathcal{X}}, \rho}(\boldsymbol{\theta})$; $i\leftarrow i+1$
            \EndWhile
            \State\Return$\boldsymbol{\theta}$
            \EndFunction
        \end{algorithmic}
\end{algorithm}
\subsubsection{Augment of Training Samples} 
\label{sub:falsification}
Here, we utilize the satisfiability modulo theories (SMT) solver \cite{gao2012deltacomplete} to analytically check if the function learned by \texttt{MinRisk} is a legitimate Lyapunov function. This is equivalent to searching for state vectors $\mathbf{x}\in\mathcal{B}_u$ that satisfy
\begin{equation} \label{eq:first_cdt}
    (V_{\boldsymbol{\theta}}(\mathbf{x})\le 0 \lor \dot{V}_{\boldsymbol{\theta}}\ge 0) \land (\norm{\mathbf{x}}_2^2 \ge l^2)
\end{equation}
where $l$ is a small scalar; and $\norm{\mathbf{x}}_2^2 \ge l^2$ is added for avoiding numerical issues of the SMT solver \cite{NIPS2019_8587}. The state vectors $\mathbf{x}\in\mathcal{B}_u$ satisfy condition \eqref{eq:first_cdt} are termed \emph{counterexamples} which can be found by the SMT solver, such as \texttt{dReal} \cite{gao2012deltacomplete}. Denote by $\mathcal{C}$ the set that consists of the counterexamples found by the SMT solver.
If $\mathcal{C}$ is not an empty set, the learned function is not a Lyapunov function and the richness of the training set $\mathcal{X}$ is enhanced by adding counterexamples in $\mathcal{C}$ to $\mathcal{X}$. The procedure of augmenting the training samples is presented in the ``\texttt{AddSample}'' function of Algorithm \ref{alg:learning_Lyapunov_function}. 

The function \texttt{LearnFunc} of Algorithm \ref{alg:learning_Lyapunov_function} summarizes the overall procedure of updating the unknown parameter $\boldsymbol{\theta}$ and augmenting the training set $\mathcal{X}$. In \texttt{LearnFunc}, $n_{\text{i}}$ is the maximum iteration times defined by users.
\begin{algorithm}
    \caption{Learning Lyapunov Function} \label{alg:learning_Lyapunov_function}
        \begin{algorithmic}[1]
            \Function{\tt{AddSample}}{$\mathcal{X},V_{\boldsymbol{\theta}},\mathbf{f}, u$}
            \State $\kappa \leftarrow 1$ 
            \State Check $\eqref{eq:first_cdt}$ in $\mathcal{B}_u$ and find $\mathcal{C}$ by \texttt{dReal}
            \If{$\mathcal{C}=\emptyset$} $\kappa \leftarrow 0$ \algorithmiccomment{No counterexamples found}
            \Else $\quad \mathcal{X}\leftarrow \mathcal{C}\cup\mathcal{X}$ \algorithmiccomment{Add counterexamples to $\mathcal{X}$}
            \EndIf
            \State\Return$\mathcal{X},\kappa$
            \EndFunction
            \Function{\tt{LearnFunc}}{$\mathcal{X},\boldsymbol{\theta}_0,\mathbf{f}, u,p,r,\eta,\alpha,\beta,\gamma,\tau, n_{\text{i}}$}
            \State $\kappa \leftarrow 1$; $j\leftarrow 0$
            \While{$(\kappa = 1)\land (j \le n_{\text{i}})$}
            \State $\boldsymbol{\theta}\leftarrow\texttt{MinRisk}(\boldsymbol{\theta}_0, \mathcal{X}, \mathbf{f}, p,r, \eta, \alpha, \beta, \gamma, \tau)$
            \State $\boldsymbol{\theta}_0\leftarrow \boldsymbol{\theta}; j \leftarrow j+r$
            \State $\mathcal{X}, \kappa \leftarrow \texttt{AddSample}(\mathcal{X},V_{\boldsymbol{\theta}},\mathbf{f}, u)$
            \EndWhile
            \If{$\kappa = 0$} $V_{\boldsymbol{\theta^*}} \leftarrow V_{\boldsymbol{\theta}} - V_{\boldsymbol{\theta}}(\mathbf{0}_m)$
            \Else $\quad V_{\boldsymbol{\theta^*}} \leftarrow \emptyset$
            \EndIf
            \State\Return$V_{\boldsymbol{\theta^*}}$
            \EndFunction
        \end{algorithmic}
\end{algorithm}

\subsection{Security Region Estimation Algorithm} 
\label{sub:SREstimation}
Given a Lyapunov function $V_{\boldsymbol{\theta^*}}$ with its valid region $\mathcal{B}_u$, we proceed to tune $d$ in \eqref{eq:S_d} so that the estimated security region is maximized. The optimal $d^*$ is determined by solving \cite{MIT}
\begin{subequations} \label{eq:d_star_original}
    \begin{align}
        &d^* = \min_{\mathbf{x}} V_{\boldsymbol{\theta}^*}(\mathbf{x}) \label{eq:cost_func}\\
        &\textrm{s.t.} \quad \norm{\mathbf{x}}_2^2 = u^2. \label{eq: ball_constrain}
    \end{align}
\end{subequations}
The state vectors satisfying the equality constrain \eqref{eq: ball_constrain} constitute the boundary of the valid region $\mathcal{B}_u$ of $V_{\boldsymbol{\theta}^*}$. Equation \eqref{eq:cost_func} essentially says that $d^*$ is the minimal value of $V_{\boldsymbol{\theta}^*}(\mathbf{x})$ evaluated along $\mathcal{B}_u$'s boundary. 

The optimization \eqref{eq:d_star} can be solved by finding \emph{critical points} defined as follows. The Lagrangian $L(\mathbf{x}, \phi)$ of \eqref{eq:d_star} is
\begin{equation}
    L(\mathbf{x},\phi) = \phi(\norm{\mathbf{x}}_2^2 - u^2)+V_{\boldsymbol{\theta}^*}(\mathbf{x}).
\end{equation}
where $\phi\in\mathbb{R}$. Define a set $\mathcal{P}$ by
\begin{equation}
    \mathcal{P}:=\left\{\mathbf{x}\in\mathbb{R}^m\Bigg| \frac{\partial L(\mathbf{x},\phi)}{\partial \mathbf{x}}=0, \norm{\mathbf{x}}_2^2 -u^2=0\right\}.
\end{equation}
Each element of the set $\mathcal{P}$ is a critical point.
The global minimum of $V_{\boldsymbol{\theta}^*}$ over $\mathcal{B}_u$'s boundary occurs at one of the critical points. 
Finding $\mathcal{P}$ is equivalent to obtaining all solutions to
\begin{equation} \label{eq:two_alg_eq}
    2\phi\mathbf{x} + \frac{\partial V_{\boldsymbol{\theta}^*}}{\partial \mathbf{x}} = \mathbf{0}_{m}; \quad \norm{\mathbf{x}}_2^2 - u^2 =0.
\end{equation}
Unknown parameters $W_1$, $W_2$, $\mathbf{b}_1$, and $\mathbf{b}_2$ in \eqref{eq:hidden_layer} and \eqref{eq:output_layer} can be updated by the $\theta^*$ returned by Algorithm \ref{alg:learning_Lyapunov_function}.
Denote by $W_1^*$, $W_2^*$, $\mathbf{b}_1^*$, and $\mathbf{b}_2^*$ the updated version of $W_1$, $W_2$, $\mathbf{b}_1$, and $\mathbf{b}_2$, respectively.
In \eqref{eq:two_alg_eq},
\begin{equation} \label{eq:algebraic_replacement}
    \frac{\partial V_{\boldsymbol{\theta}^*}}{\partial \mathbf{x}}=(1-V_{\boldsymbol{\theta}^*}(\mathbf{x})^2)W_2^*W_1^*\Lambda
\end{equation}
where $\Lambda = \text{diag}\left(1-\tanh^2(c^*_{1,1}),\ldots, 1-\tanh^2(c^*_{1,p}) \right)$, whence $[c_{1,1}^*, \ldots, c_{1,p}^*]^{\top}=W_1^*\mathbf{x}+\mathbf{b}_1^*$. With \eqref{eq:algebraic_replacement}, \eqref{eq:two_alg_eq} becomes algebraic equations whose compact form is
\begin{equation} \label{eq:cpt_form_h}
    \mathbf{h}(\mathbf{x}, \phi) = \mathbf{0}_{m+1}.
\end{equation}
The Newton-Krylov (NK) method \cite{KNOLL2004357} can solve \eqref{eq:cpt_form_h} for $\mathbf{x}$ and $\phi$ with the initial guesses $\mathbf{x}_0$ and $\phi_0$ on solutions.
If set $\mathcal{P}$ is available,
\begin{equation} \label{eq:d_star}
    d^* = \min_{\mathbf{x}\in \mathcal{P}}V_{\boldsymbol{\theta}^*}(\mathbf{x}).
\end{equation}
Then, the corresponding security region is
    \begin{equation} \label{eq:S_d_star}
    \mathcal{S}_{d^*} = \{\mathbf{x}\in\mathcal{B}_u|V_{\boldsymbol{\theta}^*}(\mathbf{x})<d^*\}.
\end{equation}

With the Lyapunov function learned by \texttt{LearnFunc}, the procedure to estimating a security region is provided by the \texttt{SREst} function of Algorithm \ref{alg:security_region_estimation}, where $\texttt{NK}(\mathbf{h},\mathbf{x}_0, \phi_0)$ denotes the procedure of solving $\mathbf{h}(\mathbf{x}, \phi) = \mathbf{0}_{m+1}$ with the initial guesses $\mathbf{x}_0$ and $\phi_0$ using the NK method; and the \texttt{NK} procedure returns $\mathbf{x}^*$ and $\phi^*$ which constitute a solution to $\mathbf{h}(\mathbf{x}, \phi) = \mathbf{0}_{m+1}$. The solution found by the \texttt{NK} procedure depends on the initial guesses $\mathbf{x}_0$ and $\phi_0$. To find all critical points, the \texttt{SREst} function repetitively solves \eqref{eq:cpt_form_h} for $n_{\text{sr}}$ times. For each time of solving \eqref{eq:cpt_form_h}, $\mathbf{x}_0$ and $\phi_0$ are randomly realized. The \texttt{Main} function of Algorithm \eqref{alg:security_region_estimation} summarizes the procedure described in Sections \ref{sub:asymptotic_stability_check_and_security_region_estimation}, \ref{sub:learning_lyapunov_function}, and \ref{sub:SREstimation}. Note that checking asymptotic stability of the given equilibrium (Lines 14-16 of Algorithm \ref{alg:security_region_estimation}) is a prerequisite for learning a Lyapunov function and estimating an optimal security region.
\begin{algorithm}
    \caption{Security Region Estimation} \label{alg:security_region_estimation}
        \begin{algorithmic}[1]
            \Function{\tt{SREst}}{$V_{\boldsymbol{\theta}^*}, u, n_{\text{sr}}$}
            \State $\mathcal{P}\leftarrow \emptyset$; construct $\mathbf{h}$ by \eqref{eq:two_alg_eq}, \eqref{eq:algebraic_replacement}
            \For{$k = 1, 2, \ldots, n_{\text{sr}}$}
            \State Pick a random $\mathbf{x}_0$ in $\{\mathbf{x}_0\in\mathbb{R}^m|\norm{\mathbf{x}_0}_2^2 = u^2\}$
            \State Pick a random $\phi_0 \in \mathbb{R}$
            \State $\mathbf{x}^*, \phi^* \leftarrow\texttt{NK}(\mathbf{h}, \mathbf{x}_0, \phi_0)$
            \If{$\mathbf{x}^*\notin \mathcal{P}$} $\mathcal{P}\leftarrow \mathcal{P}\cup \mathbf{x}^*$
            \EndIf
            \EndFor
            \State Obtain $S_{d^*}$ via \eqref{eq:d_star}, \eqref{eq:S_d_star}
            \State\Return $S_{d^*}$
            \EndFunction
            \Function{\tt{Main}}{$\mathbf{f},u, p, q, \boldsymbol{\theta}_0, r, \eta, \alpha, \beta, \gamma, \tau, n_{\text{sr}}, n_{\text{i}}$}
            \State Linearize $\mathbf{f}$ to obtain $A$ in \eqref{eq:Ax}
            \State Compute eigenvalues $\lambda_i$ of $A$ $\forall i = 1,2,\ldots m$
            \If{\eqref{eq:stability_check} holds} \algorithmiccomment{Asymptotic stability check}
            \State Construct $\mathcal{X}$ by randomly picking $q$ vectors in $\mathcal{B}_u$
            \State $V_{\boldsymbol{\theta}^*}\leftarrow \texttt{LearnFunc}(\mathcal{X},\boldsymbol{\theta}_0,\mathbf{f}, u,p, r,\eta,\alpha,\beta,\gamma,\tau, n_{\text{i}})$
            \If{$V_{\boldsymbol{\theta}^*} \ne\emptyset$}
            \State $S_{d^*}\leftarrow\texttt{SREst}(V_{\boldsymbol{\theta}^*}, u, n_{\text{sr}})$
            \State\Return $V_{\boldsymbol{\theta}^*}$, $S_{d^*}$
            \Else \quad Request for tunning user-defined parameters
            \EndIf
            \Else \quad Request for tuning parameters in \eqref{eq:f_x}
            \EndIf
            \EndFunction
        \end{algorithmic}
\end{algorithm}

\subsection{Parameter Tuning} 
\label{sub:parameter_tunning}

In Algorithm \ref{alg:security_region_estimation}, the empirical settings of $\boldsymbol{\theta}_0, p, q, r, \eta$, $n_{\text{sr}}$, $n_{\text{i}}$ and $\tau$ are provided as follows: the random initial guess $\boldsymbol{\theta}_0$ is obtained by the initialization procedure reported in \cite{7410480}; $p\ge 2m$; $q$, $n_{\text{sr}}$, and $n_{\text{i}}$ are $500$, $100$, and $5000$, respectively; integer $r\in[10,30]$; $\tau\in[0.1,0.5]$; and $\eta \in[0.01, 0.02]$.

\begin{figure}[b]
        \centering
        \subfloat[]{\includegraphics[width=1.5in]{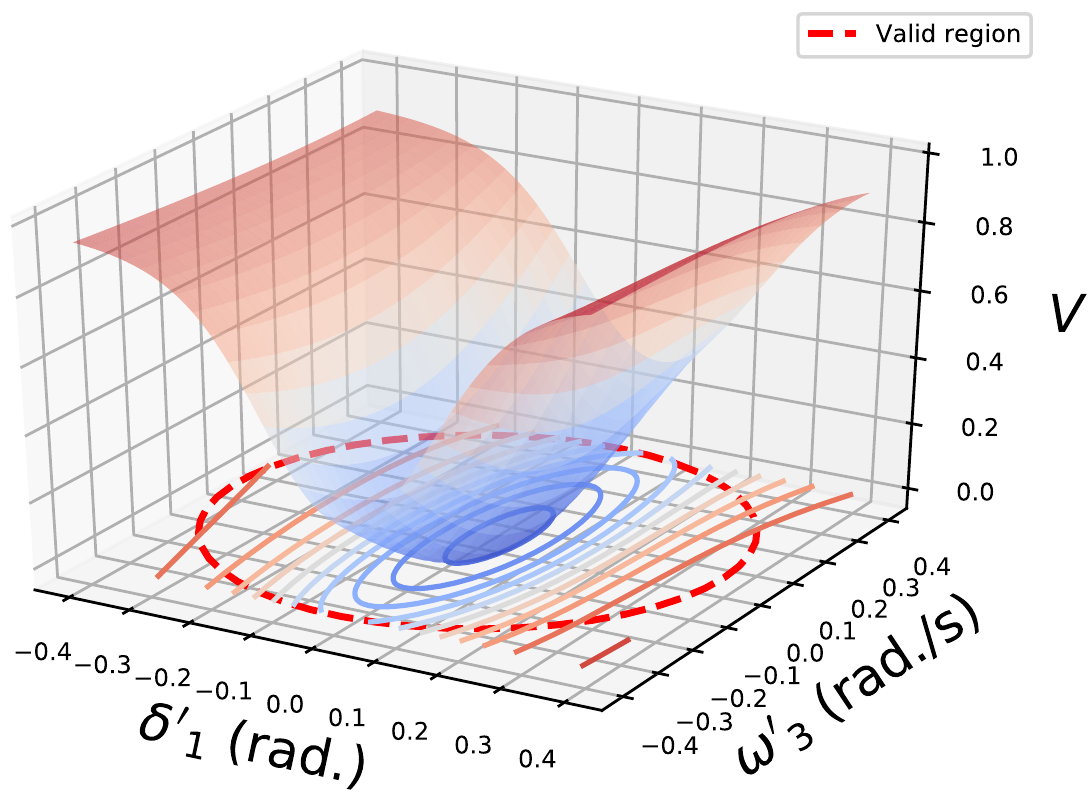}}
        \hfil
        \subfloat[]{\includegraphics[width=1.5in]{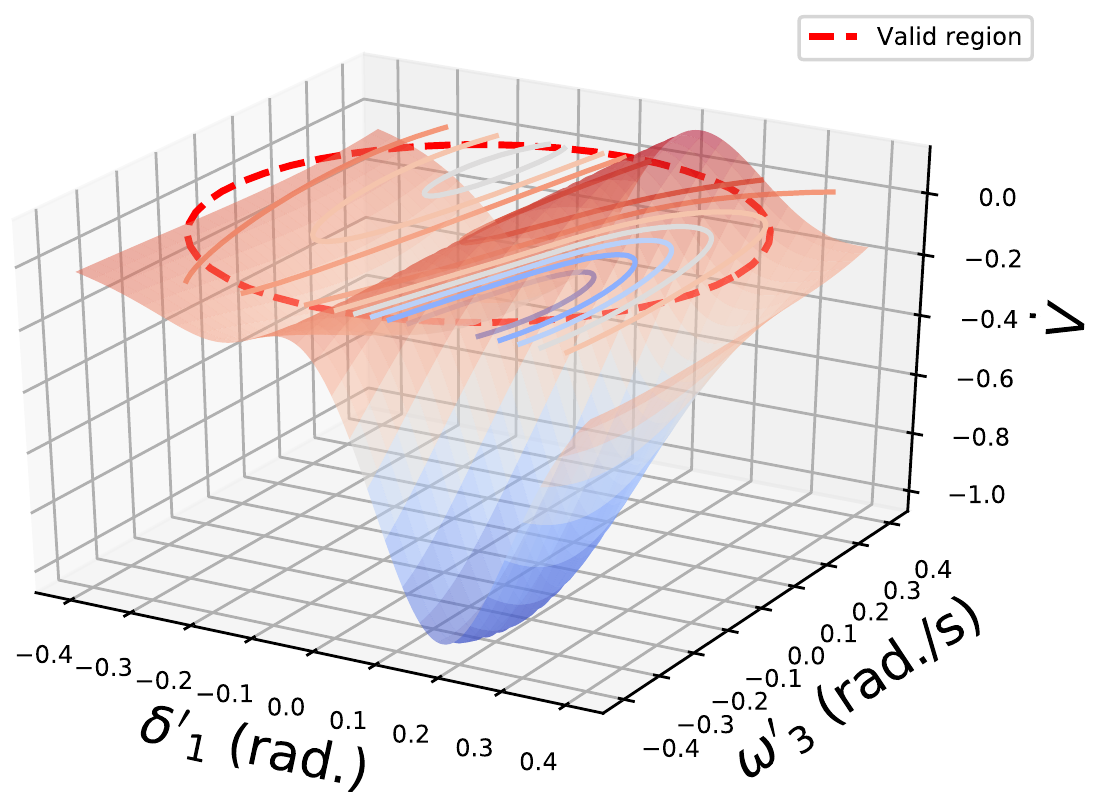}}
        \hfil
        \caption{Visualization of the function (a) and its time derivative (b) after $n_{\text{i}}$ times of parameter update: the function is NOT a Lyapunov function.}
        \label{fig:LF_bad_example}
    \end{figure}
Given a set of user-defined parameters, it is possible that the \texttt{Main} function returns an empty set $\emptyset$, meaning that the function fails to find a Lyapunov function valid in $\mathcal{B}_u$ within $n_{\text{i}}$ iterations. Solutions to such a situation include 1) decreasing $u$; 2) changing $\boldsymbol{\theta}_0$; and 3) tunning $\alpha$, $\beta$, and $\gamma$. Solution 1 works because there may not be a Lyapunov function in a large ball. Solving \eqref{eq:optimization} using gradient-based methods depends on the initial guess on the solution, which justifies Solution 2. 

Next we present an empirical procedure to tune $\alpha$, $\beta$, and $\gamma$. Denote by $\boldsymbol{\theta}_{n_{\text{i}}}$ the $n_{\text{i}}$-th update of $\boldsymbol{\theta}$ in \texttt{LearnFunc}. The function $V_{\boldsymbol{\theta}_{n_{\text{i}}}}$ and its time derivative $\dot{V}_{\boldsymbol{\theta}_{n_{\text{i}}}}$ can be visualized in subspace of $\mathcal{B}_u$. The visualization may suggest which condition(s) in Definition \ref{def:Lyapunov_function} is (are) violated, thereby pointing out the ``direction'' of tunning $\alpha$, $\beta$, and $\gamma$. For example, suppose that one needs to learn a Lyapunov function for a system whose state variables are $[\delta_1', \delta_2', \delta_3', \omega_3']$ using \texttt{LearnFunc}. After $n_{\text{i}}$-time parameter updates, the function with parameter $\boldsymbol{\theta}_{n_{\text{i}}}$ and its time derivative can be visualized by numerically evaluating the functions within $\mathcal{B}_{u}$'s projection to the $\delta_1'$-$\omega_3'$ plane with $\delta_2'=\delta_3'=0$. Suppose that the visualization is given in Figure \ref{fig:LF_bad_example}. As shown in Figure \ref{fig:LF_bad_example}, the function with parameter $\boldsymbol{\theta}_{n_{\text{i}}}$ is \emph{not} a Lyapunov function in $\mathcal{B}_{0.4}$, because its time derivative is not negative in $\mathcal{B}_{0.4}$, although the function is positive. Figure \ref{fig:LF_bad_example} indicates that with other parameters fixed, one may need to increase the penalty resulting from the violation of the second condition of Definition \ref{def:Lyapunov_function}, i.e., increasing $\beta$ in \eqref{eq:Lyapunov_risk}.



\section{Numerical Experiments} 
This section tests and validates the proposed method in a grid-connected microgrid, a three-microgrid interconnection with mixed dynamics, and the IEEE 123-node feeder. 
\label{sec:numerical_experiments}
\subsection{A Grid-connected Microgrid} 
\begin{figure}
    \centering
    \includegraphics[width = 2.5in]{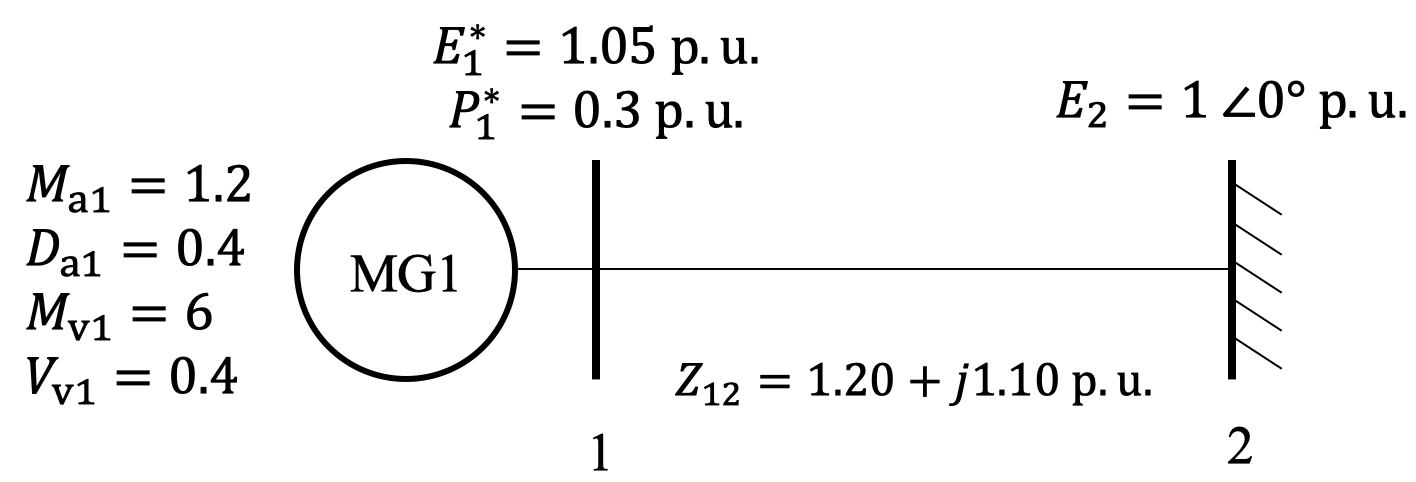}
    \caption{A grid-connected microgrid \cite{HICSS2021Huang}}
    \label{fig:onMG}
\end{figure}

Figure \ref{fig:onMG} shows a grid-connected microgrid (MG) with angle-droop control. 
The user-defined parameters required in Algorithm \ref{alg:security_region_estimation} are listed in Table \ref{tab:alg_para}.
\label{sub:a_grid_connected_microgrid}
\begin{table}
    \caption{User-defined Parameters of Algorithm \ref{alg:security_region_estimation}}
    \label{tab:alg_para}
    \centering

    \begin{tabular}{c|c|c|c|c|c|c}
    \hline

    \hline
    \textbf{Case Name} & $p$ & $q$ & $n_{\text{sr}}$ & $n_{\text{i}}$ & $r$ &$\tau$\\
    \hline
         A Grid-connected Microgrid& $6$& $500$ & $100$ & $5000$ & $10$ & $0.1$\\
    \hline
         Three Networked Microgrids &$8$& $500$ &$100$ & $5000$ & $30$ & $0.1$\\
    \hline
         IEEE 123-node Feeder       &$8$& $500$   &$100$   &$5000$    &$10$ &$0.5$\\
    \hline

    \hline
    \textbf{Case Name} & $\eta$ &$u$& $\alpha$ &$\beta$ & $\gamma$ & N/A\\
    \hline
         A Grid-connected Microgrid& $0.01$ & $1.5$ &$1$ &$5$&$0$& N/A\\
         \hline
         Three Networked Microgrids& $0.02$ & $0.4$ &$3$ &$1$ &$3$&N/A\\
         \hline
         IEEE 123-node Feeder& $0.01$       & $0.7$ &$1$ &$1$ &$0$&N/A\\
        \hline

        \hline

    \end{tabular}
\end{table}

\subsubsection{Learned Lyapunov Function} 
\label{ssub:lyapunov_function_learned_Onemac}
\begin{figure}
        \centering
        \subfloat[]{\includegraphics[width=1.65in]{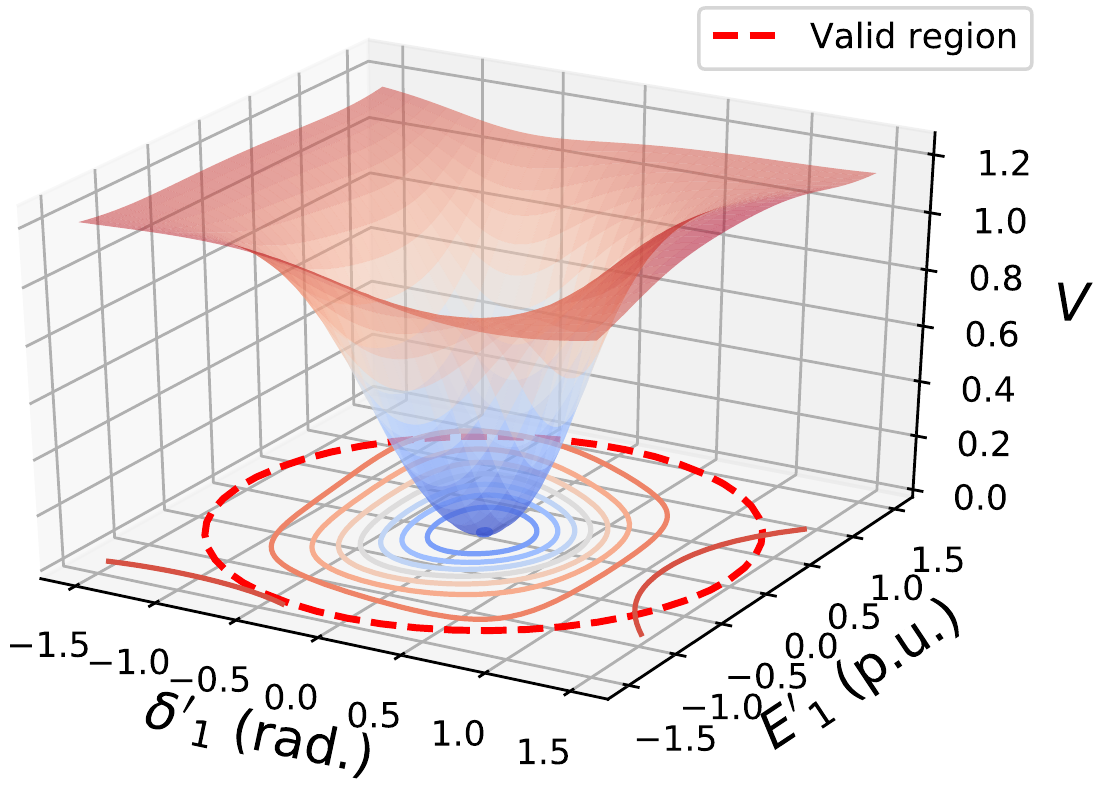}}
        \hfil
        \subfloat[]{\includegraphics[width=1.65in]{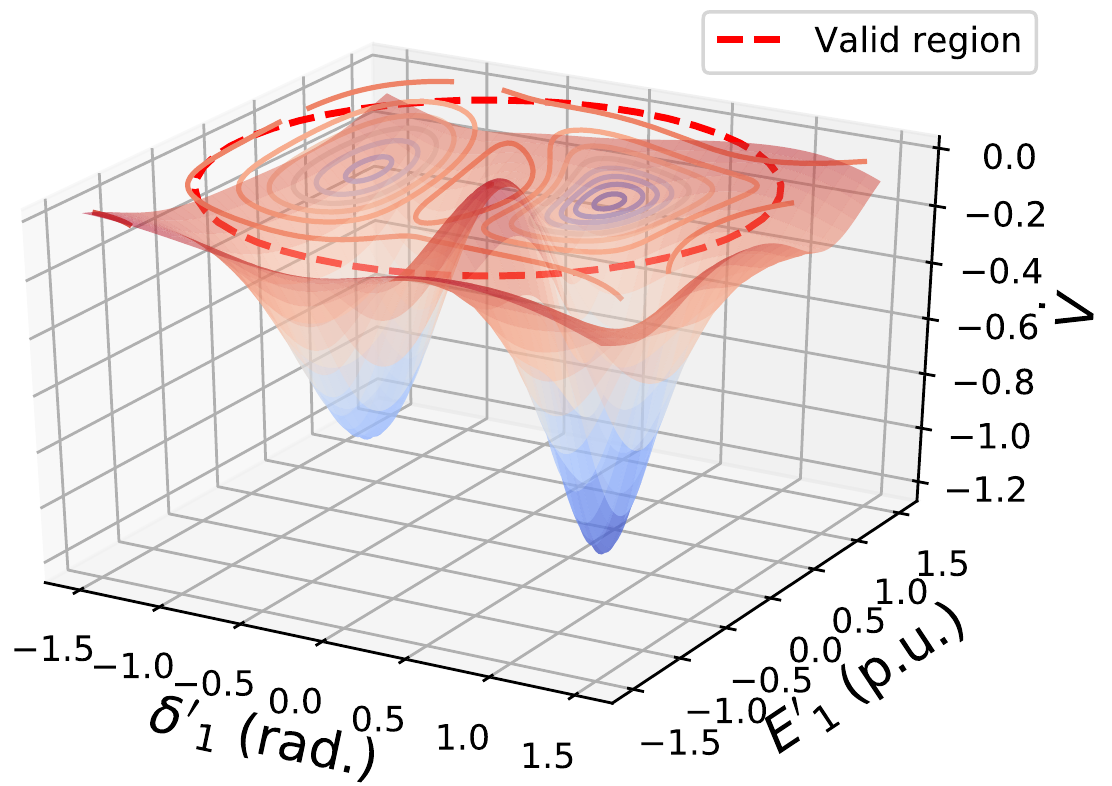}}
        \hfil
        \caption{(a) Lyapunov function and (b) its time derivative for a grid-tied MG}
        \label{fig:LF_OneMG}
    \end{figure}

After $500$ times of parameter updates, which takes $32.18$ seconds, Algorithm \ref{alg:learning_Lyapunov_function} outputs a Lyapunov function. Figure \ref{fig:LF_OneMG} visualizes the Lyapunov function learned and its time derivative. As shown in Figure \ref{fig:LF_OneMG}, the function learned is positive definite in the valid region $\mathcal{B}_{1.5}$ and its time derivative is negative definite in $\mathcal{B}_{1.5}$. This suggests that the function learned is a Lyapunov function in $\mathcal{B}_{1.5}$.

\subsubsection{Estimated Security Region} 
\label{ssub:security_region_estimated_OneMac}
\begin{figure}
    \centering
    \centering
        \subfloat[]{\includegraphics[width=1.65in]{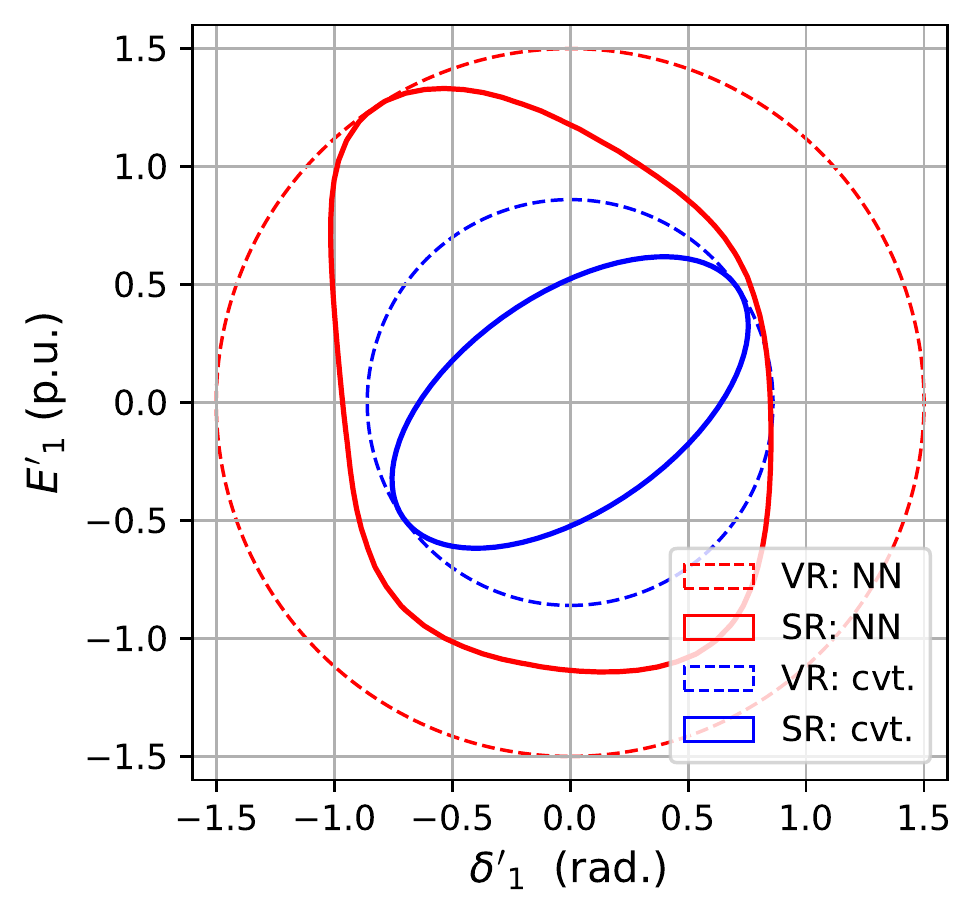}}
        \hfil
        \subfloat[]{\includegraphics[width=1.65in]{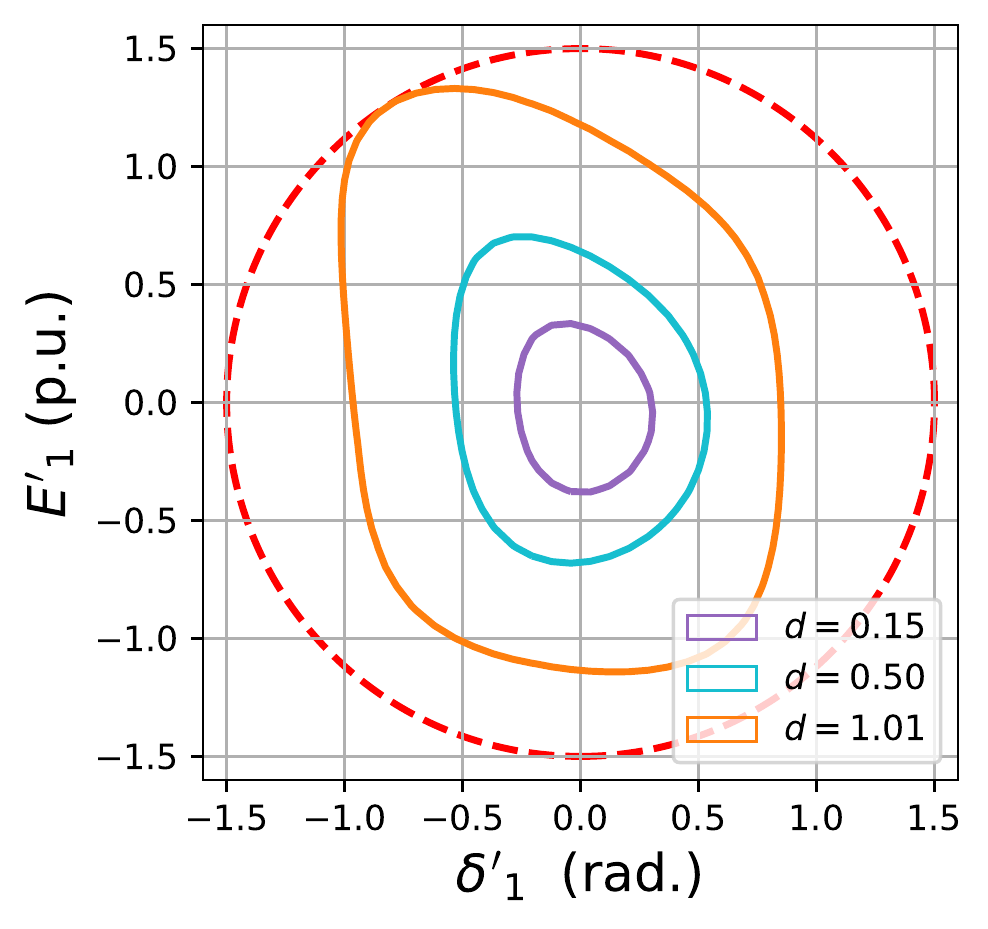}}
        \hfil
    \caption{(a) Comparison between the proposed (NN) and conventional (cvt.) methods: security region (SR) and valid region (VR). (b) An alternative way to find $\mathcal{S}_{d^*}$ by tuning $d$.}
    \label{fig:OneMG_CMP}
\end{figure}

Given the Lyapunov function learned with its valid region $\mathcal{B}_{1.5}$, the security region estimated by Algorithm \ref{alg:security_region_estimation} is $\mathcal{S}_{1.01}$ which is defined by \eqref{eq:S_d_star}. In Figure \ref{fig:OneMG_CMP}, the red-solid circle is the boundary of $\mathcal{S}_{1.01}$, while the red-dash circle is the boundary of $\mathcal{B}_{1.5}$; and the region enclosed by the red-solid circle is a security region. Besides, the \texttt{SREst} function suggests that $d^*$ in \eqref{eq:d_star_original} is $1.01$ which is attained when $\delta_1'=-0.82$ and $E_1'=1.26$.

We proceed to check the correctness of the estimated security region $\mathcal{S}_{1.01}$. Since the test system only has two state variables, given the Lyapunov function learned, the largest security region can be found without solving optimization \eqref{eq:d_star_original}. For example, we can visualize a security region $\mathcal{S}_d$ with a small $d$, say, $d=0.15$. Figure \ref{fig:OneMG_CMP}-(b) visualize $\mathcal{S}_{0.15}$. We keep increasing $d$ gradually until the boundary of $\mathcal{S}_d$ touches the boundary of $\mathcal{B}_{1.5}$ for the first time. As can be observed in Figure \ref{fig:OneMG_CMP}-(b), when $d=1.01$, the boundaries of $\mathcal{S}_d$ and $\mathcal{B}_{1.5}$ touch with each other at $(-0.82, 1.26)$. Therefore, $\mathcal{S}_{1.01}$ is the largest security region that can be estimated based on the learned Lyapunov function. The security region obtained by such a procedure is consistent with the one estimated by function \texttt{SREst}.


\subsubsection{Comparison} 
\label{ssub:comparison_study_OneMac}
The proposed method is compared with a conventional method reported in \cite{41298}. Denote by $\mathcal{S'}$ the security region estimated based on a quadratic Lyapunov function constructed in \cite{41298}. In Figure \ref{fig:OneMG_CMP}, the region enclosed by the blue-solid circle is $\mathcal{S'}$, while the blue-dash circle is the boundary of the valid region of the quadratic Lyapunov function. It can be observed that $\mathcal{S}_{1.01}$ is larger than $\mathcal{S}'$. This suggests that the propose method can provide a less conservative characterization of the security region than the conventional method.

\begin{figure}[b]
        \centering
        \subfloat[]{\includegraphics[width=1.5in]{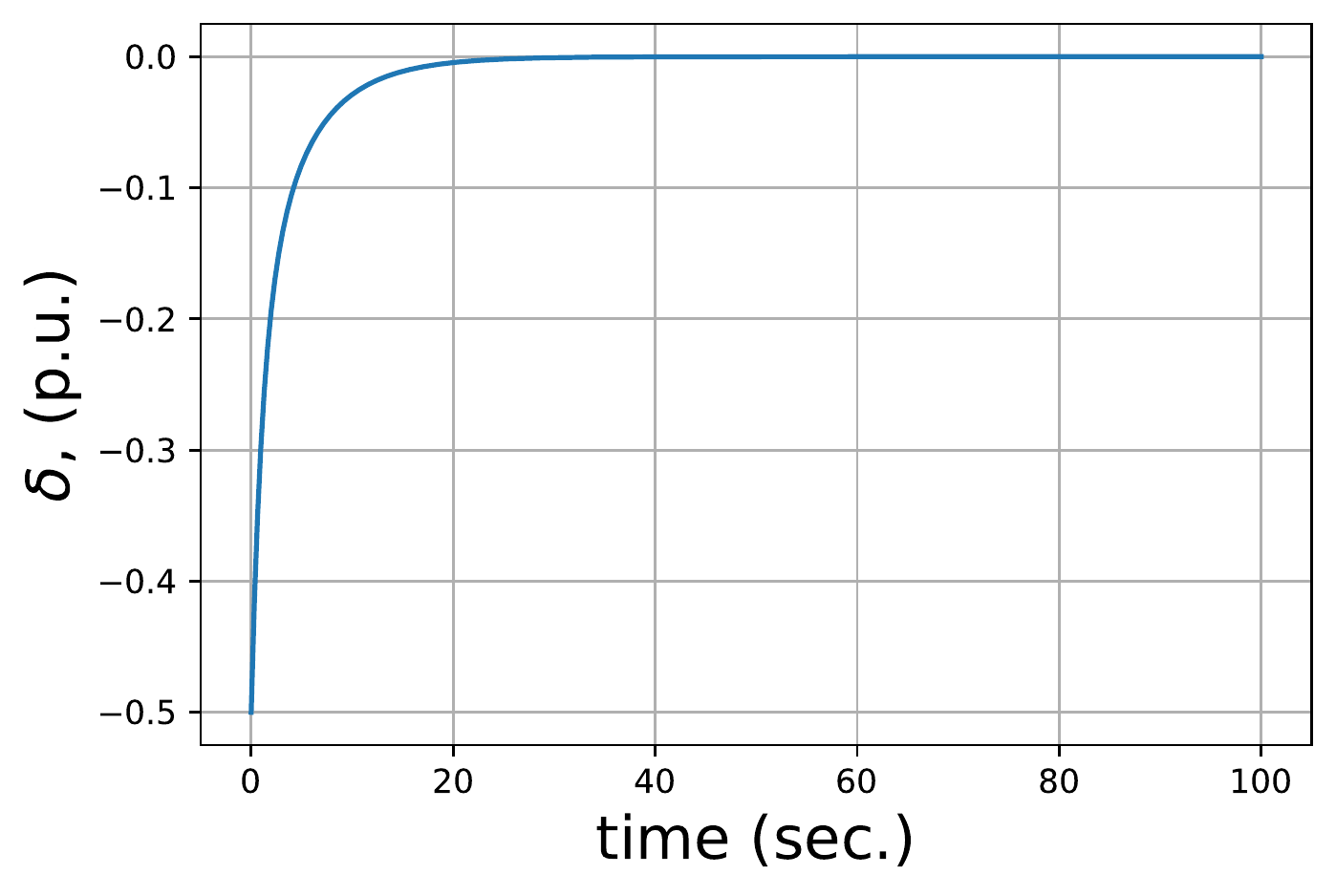}}
        \hfil
        \subfloat[]{\includegraphics[width=1.5in]{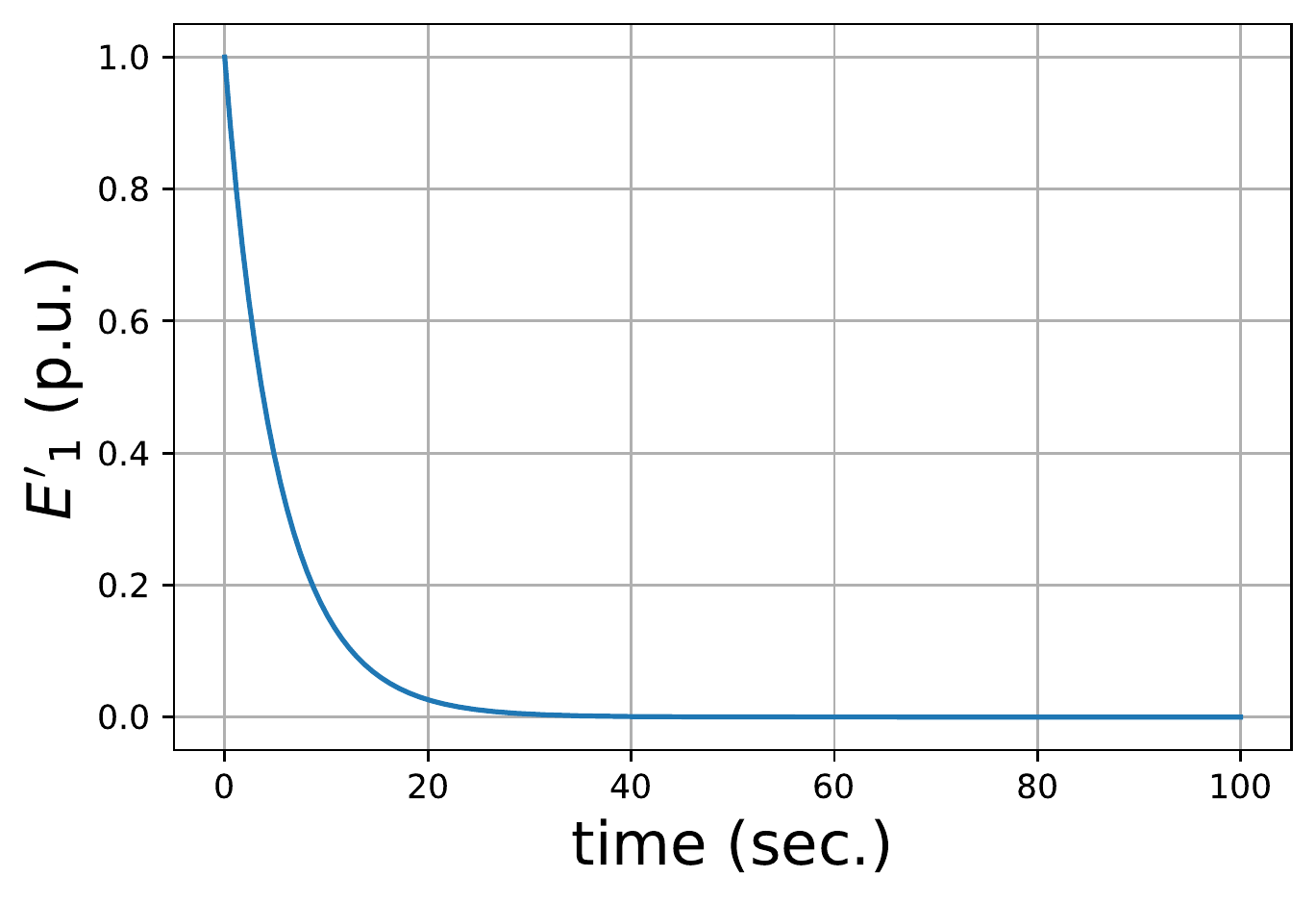}}
        \hfil
        \caption{Time-domain simulation for the grid-connected MG with initial conditions $\delta_1'(0)=-0.5$ rad. and $E'_1(0) = 1$ p.u.}
        \label{fig:OneMGTimeDomain}
    \end{figure}

Suppose that the grid-connected MG has an initial condition $\mathbf{x}(0)=[-0.5, 1]^{\top}$, due to a disturbance. Such an initial condition is inside $\mathcal{S}_{1.01}$, but outside $\mathcal{S}'$. Therefore, \emph{$\mathcal{S}_{1.01}$ can conclude that the system trajectory tends to its equilibrium point, whereas $\mathcal{S}'$ can conclude nothing about the system's asymptotic behavior under such a disturbance.} The time-domain simulation shown in Figure \ref{fig:OneMGTimeDomain} confirms that all state variables tend to their pre-dispatched steady-state values.


\subsection{Three Networked Microgrids with Mixed Dynamics} 
\begin{figure}
    \centering
    \includegraphics[width = 2.5in]{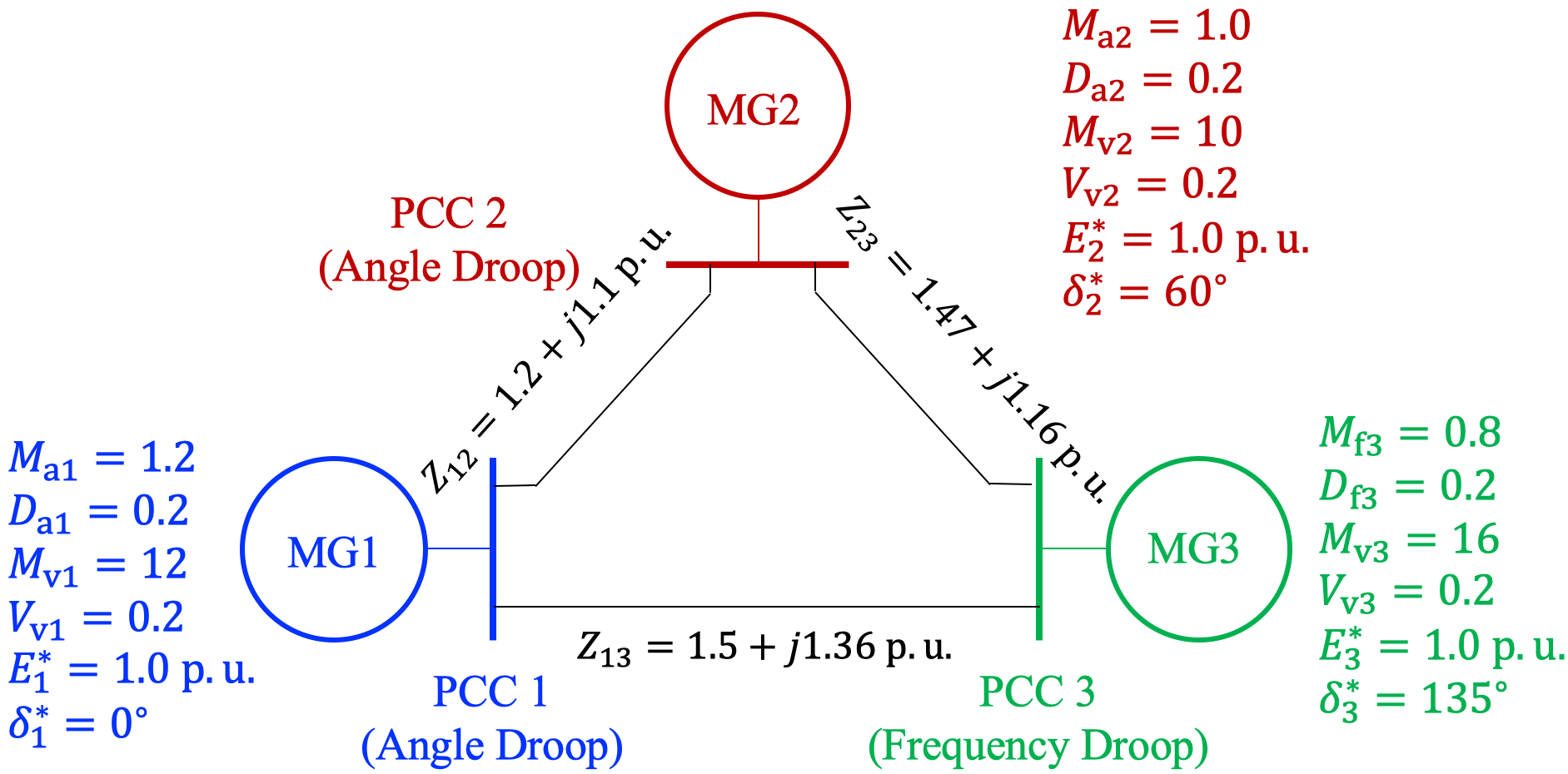}
    \caption{Three Networked Microgrids with Mixed Dynamics}
    \label{fig:ThreeMG}
\end{figure}

Figure \ref{fig:ThreeMG} shows a three-MG interconnection with mixed interface dynamics: the angle droop control is deployed in the PE interfaces of MGs 1 and 2, whereas the frequency droop control is deployed in the PE interfaces of MG 3. Since $M_{\text{v}k}\gg M_{\text{a}k}$ for $k=1,2$ and $M_{\text{v}3}\gg M_{\text{f}3}$, the time-scale separation is assumed \cite{ZhangLarge}. We focus on the asymptotic behavior of phase angle and frequency. The user-defined parameters of Algorithm \ref{alg:security_region_estimation} are listed in Table \ref{tab:alg_para}.
\label{sub:three_networked_microgrids_with_mixed_dynamics}





\subsubsection{Learned Lyapunov Function} 
\label{subsub:learned_lyapunov_function_3Mac}
\begin{figure}
        \centering
        \subfloat[]{\includegraphics[width=1.45in]{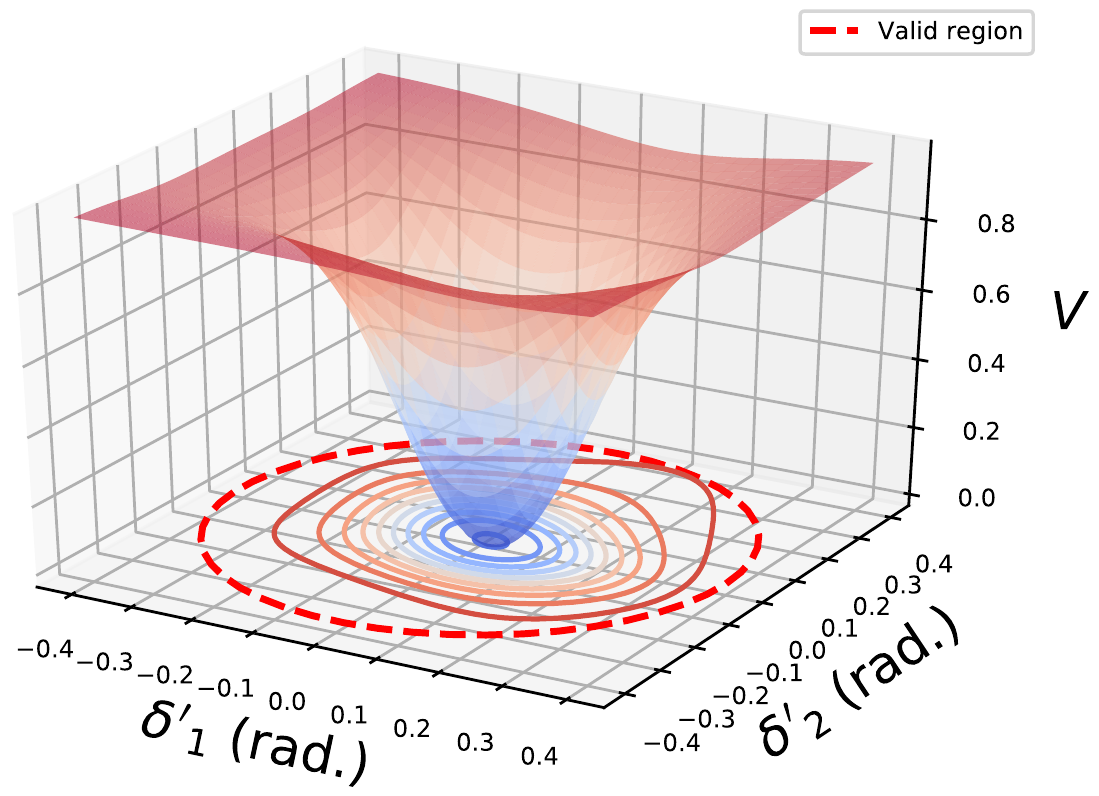}}
        \hfil
        \subfloat[]{\includegraphics[width=1.45in]{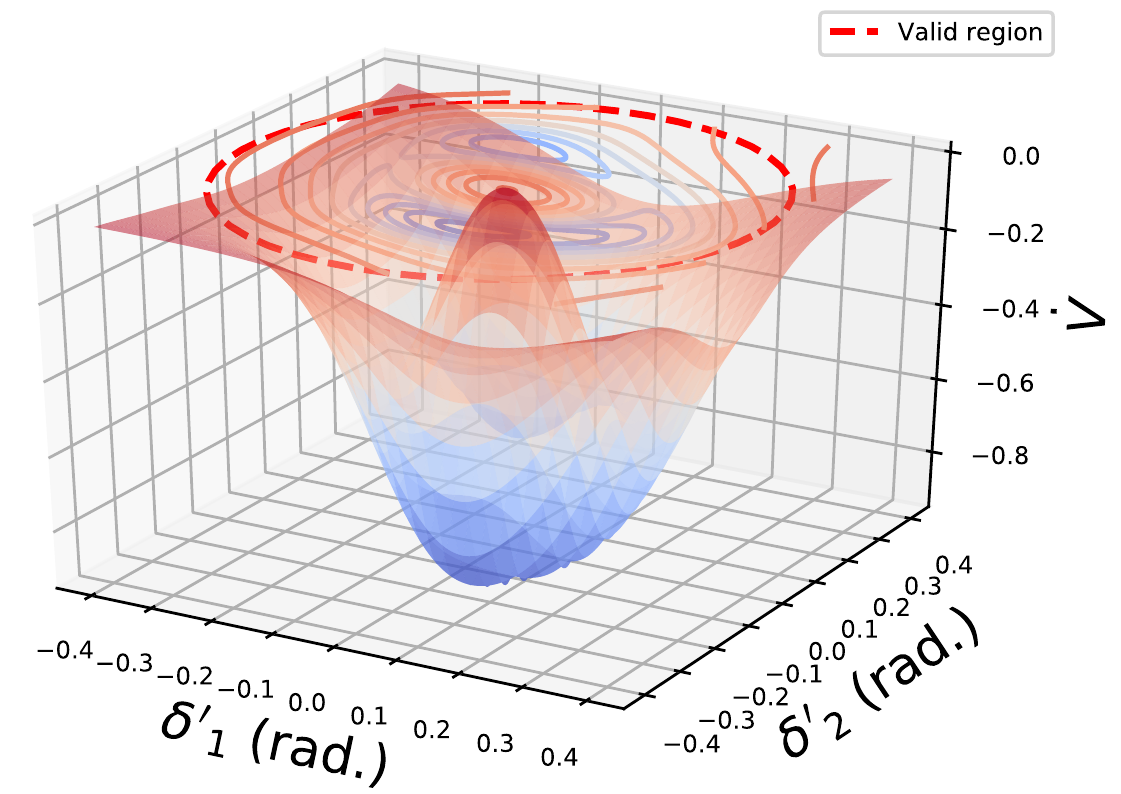}}
        \hfil
        \caption{(a) Lyapunov function and (b) time derivative for 3 networked MGs}
        \label{fig:ThreeMG_LF}
    \end{figure}

In this expertiment, after $2790$ times of parameter updates, which takes $23737.53$ seconds, Algorithm \ref{alg:security_region_estimation} outputs a Lyapunov function $V_{\boldsymbol{\theta^*}}$ valid in $\mathcal{B}_{0.4}$. Given $\delta_3'=0$ and $\omega_3'=0$, $V_{\boldsymbol{\theta^*}}$ and $\dot{V}_{\boldsymbol{\theta^*}}$ are visualized in Figure \ref{fig:ThreeMG_LF} where it is observed that $V_{\boldsymbol{\theta^*}}>0$ and $\dot{V}_{\boldsymbol{\theta^*}}<0$ in $\mathcal{B}_{0.4}$, suggesting $V_{\boldsymbol{\theta^*}}$ behaves like a Lyapunov function.


\subsubsection{Estimated Security Region} 
\label{ssub:estimated_security_region_3Mac}
\begin{figure}
        \centering
        \subfloat[]{\includegraphics[width=1.55in]{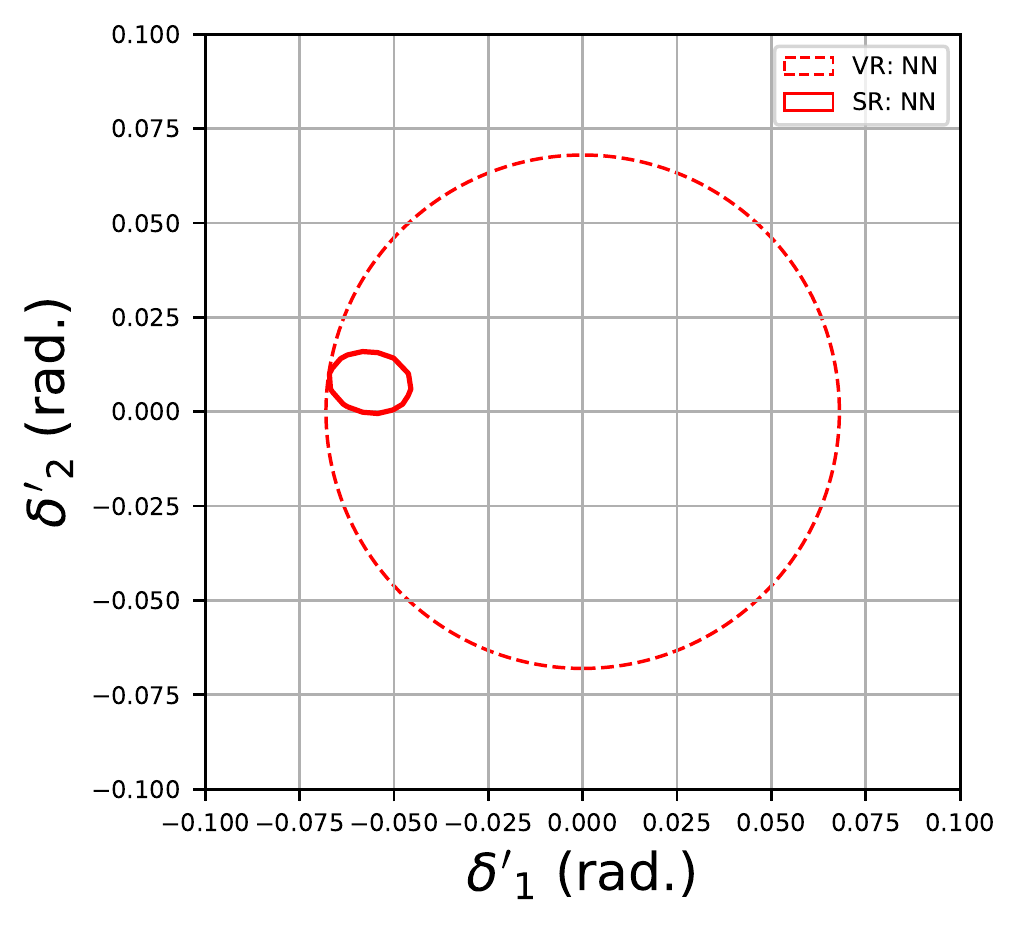}}
        \hfil
        \subfloat[]{\includegraphics[width=1.45in]{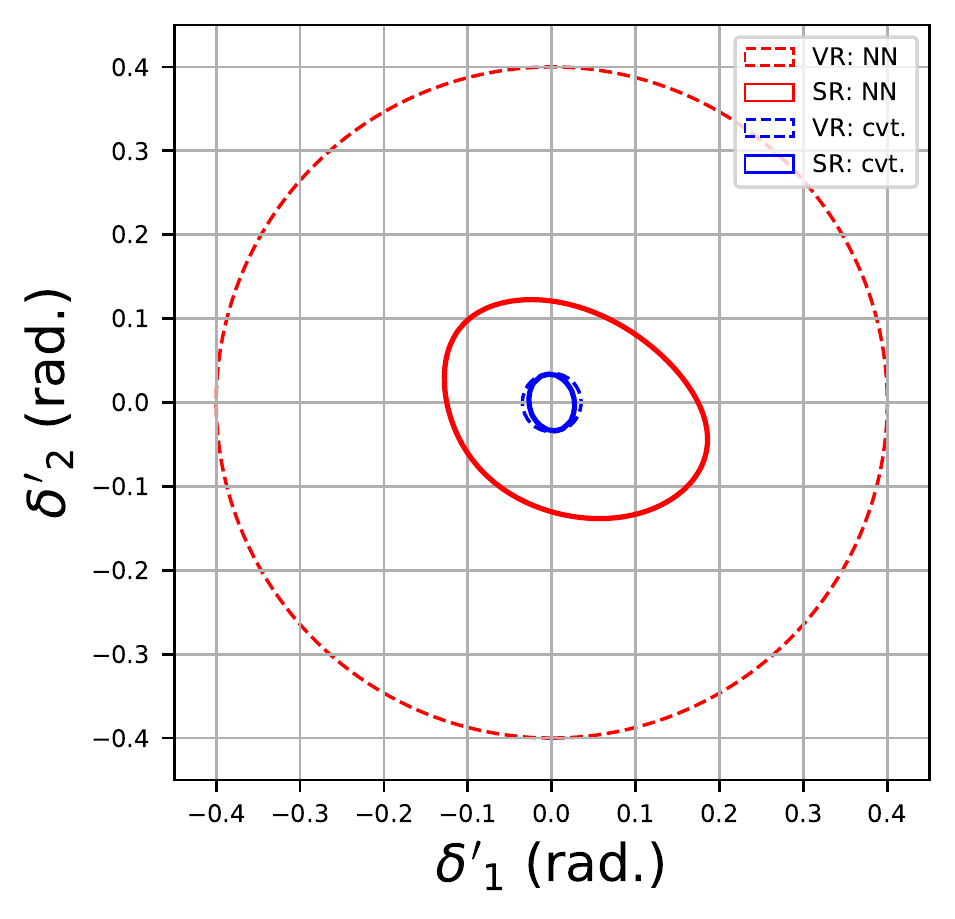}}
        \hfil
        \caption{(a) Security region (SR) and valid region (VR) around the touching point. (b) Comparison between the proposed (NN) and conventional (cvt.) methods.}
        \label{fig:ThreeMG_Comparison}
\end{figure}

With the learned Lyapunov function, Algorithm \ref{alg:security_region_estimation} provides an estimated security region $\mathcal{S}_{0.37}$. 
Figure \ref{fig:ThreeMG_Comparison}-(a) visualizes $\mathcal{S}_{0.37}$ and $\mathcal{B}_{0.4}$ in the $\delta_1'$-$\delta_2'$ space with $\delta_3'=0.37$ and $\omega_3'=-0.14$, where the red-solid circle is the boundary of $\mathcal{S}_{0.37}$, and the red-dash circle is the boundary of $\mathcal{B}_{0.4}$. The \texttt{SREst} function suggests that $d^*$ in \eqref{eq:d_star_original} is $0.37$ which is attained when $\mathbf{x}$ is $[-0.07,0.01, 0.37,-0.14]^{\top}$. Figure \ref{fig:ThreeMG_Comparison}-(a) shows that the boundary of $\mathcal{S}_{0.37}$ touches the boundary of $\mathcal{B}_{0.4}$ at point $(-0.07,0.01)$. 
\subsubsection{Comparison} 
\label{ssub:comparison_study_3Mac}
Denote by $\mathcal{S}''$ the security region estimated based on the Lyapunov function proposed in \cite{41298}. The blue-solid circle in Figure \ref{fig:ThreeMG_Comparison}-(b) represents the boundary of $\mathcal{S}''$ in the $\delta_1'$-$\delta_2'$ plane, given $\delta_3'=\omega_3'=0$. Suppose that the pre-event condition $\mathbf{x}(0)$ is $[0.1, -0.1, 0,0]^{\top}$. Since $\mathbf{x}(0)$ is inside $\mathcal{S}_{0.37}$ but outside $\mathcal{S}''$, one can conclude that \emph{all states tend to the equilibrium based on $\mathcal{S}_{0.37}$, while the asymptotic behavior the system cannot be assessed by $\mathcal{S}''$ with $\mathbf{x}(0)$.} The time-domain simulation confirms that all state variables indeed converge to their post-event steady-state values.
\begin{figure}
        \centering
        \subfloat[]{\includegraphics[width=1.55in]{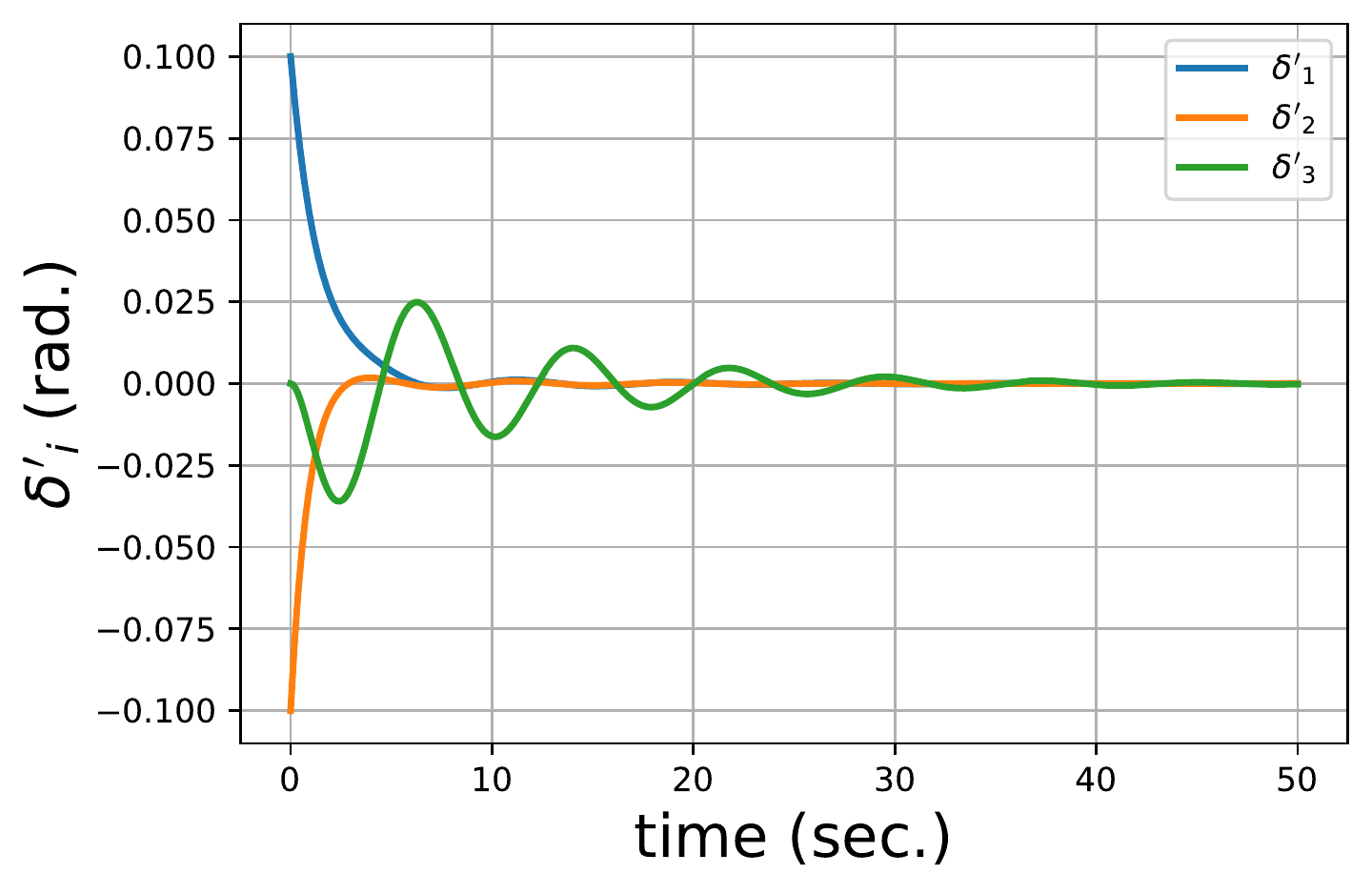}}
        \hfil
        \subfloat[]{\includegraphics[width=1.55in]{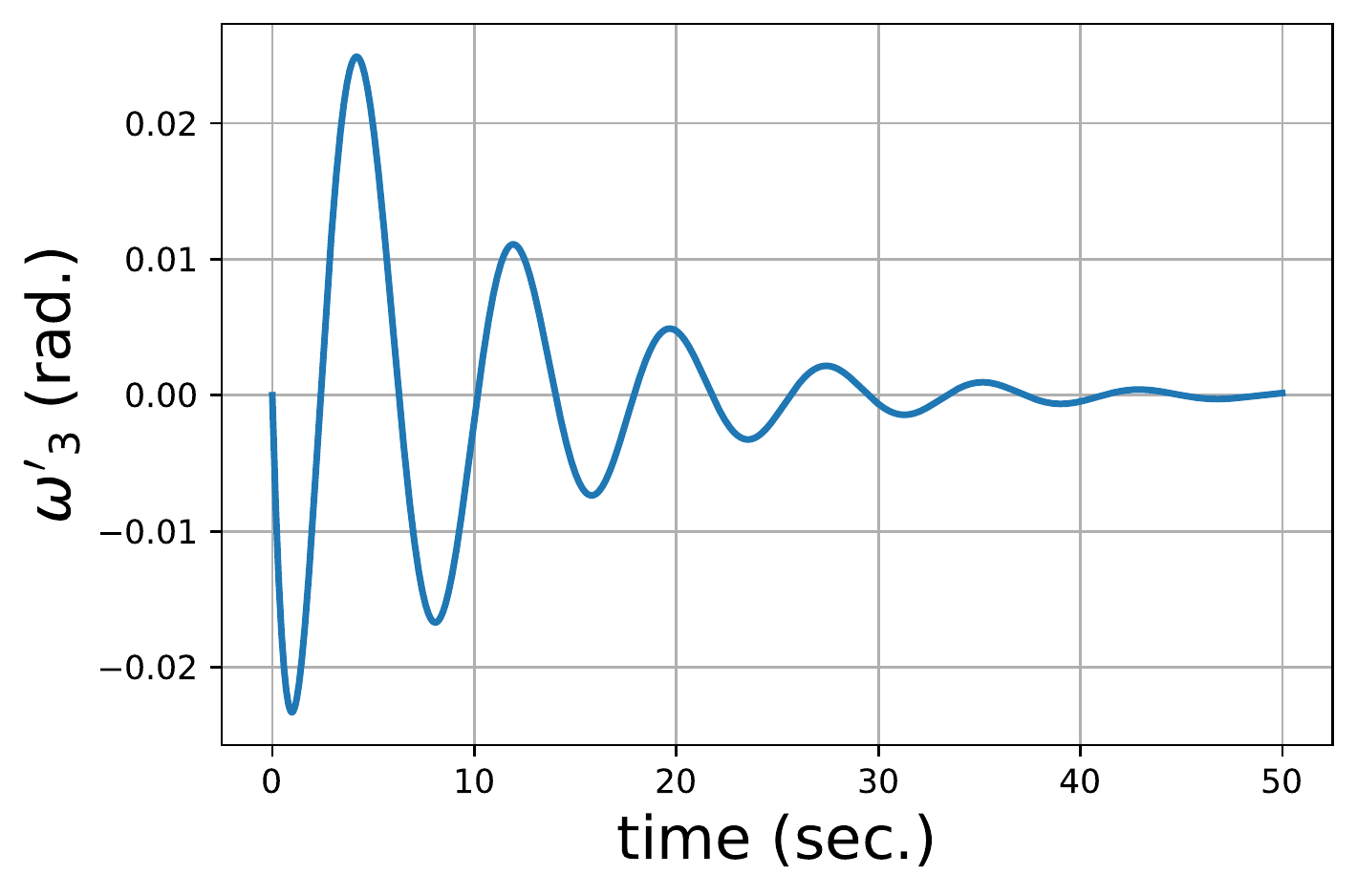}}
        \hfil
        \caption{Time-domain simulation of the 3 networked MGs with $\mathbf{x}(0)=[0.1, -0.1, 0,0]^{\top}$: (a) angle deviation and (b) frequency deviation.}
        \label{fig:ThreeMG_Time_domain}
\end{figure}

\subsection{IEEE 123-node Test Feeder} 
\label{sub:ieee_123_bus_feeder}
\begin{figure}
    \centering
    \includegraphics[width = 2in]{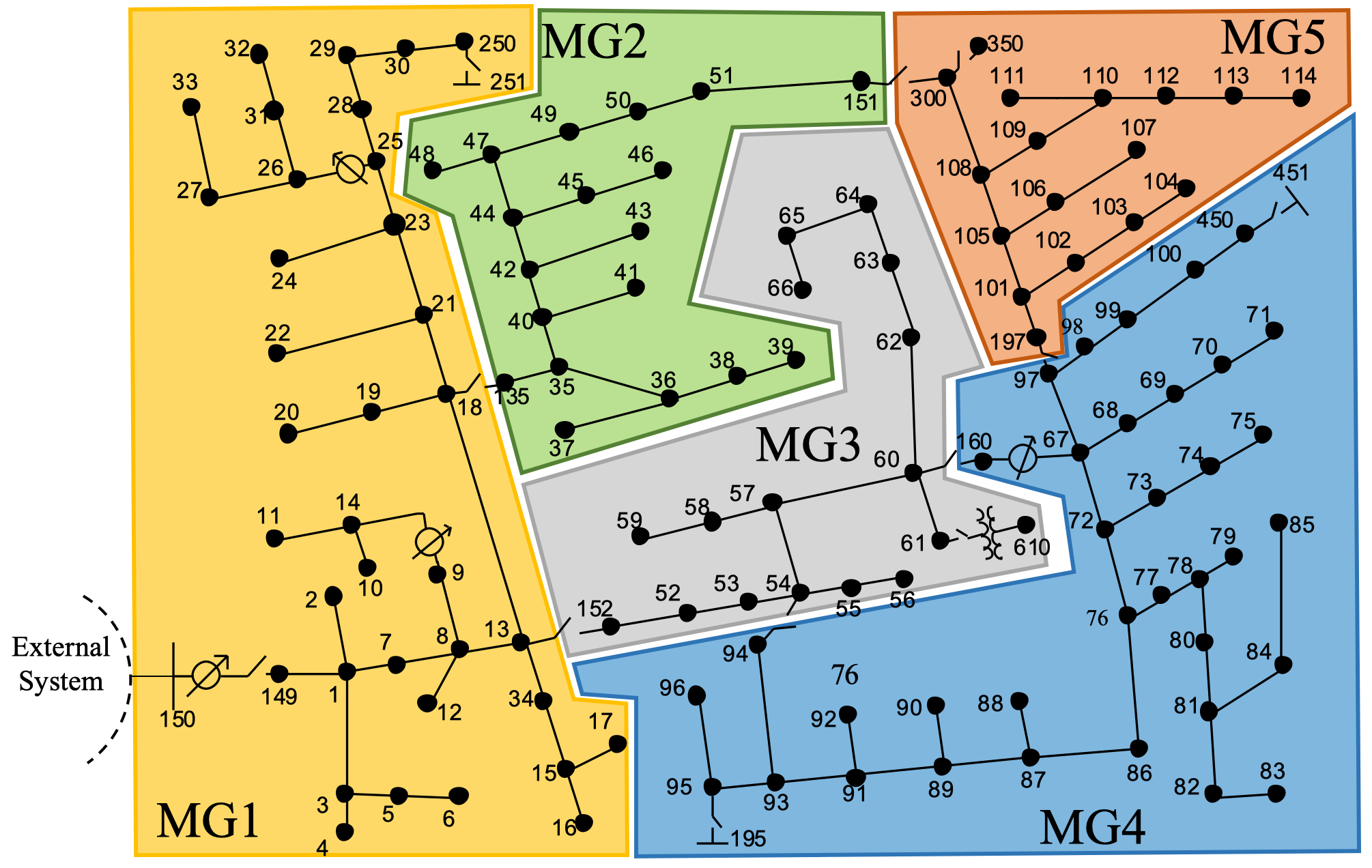}
    \caption{IEEE 123-node Test Feeder \cite{ZhangLarge}}
    \label{fig:Node123}
\end{figure}

Figure \ref{fig:Node123} shows a $123$-node distribution system \cite{8063903} which is partitioned into $5$ networked MGs \cite{ZhangLarge}. We assume that each MG is managed by its MGCC and connects to the grid via a PE interface with angle droop control \cite{ZhangLarge}. The impedances of the interconnection distribution lines are reported in Table \ref{tab:123node_line_para}. The control parameters and pre-dispatched setpoints are listed in Table \ref{tab:123_ctrl_para}. The user-defined parameters in Algorithm \ref{alg:security_region_estimation} are reported in Table \ref{tab:alg_para}. Note that the time-scale separation is assumed, as $M_{\text{v}k}\gg M_{\text{a}k}$ for $k=1,2,\ldots,5$ in Table \ref{tab:123_ctrl_para}.
\begin{table}[tbh]
    \caption{Distribution Line Parameters}
    \label{tab:123node_line_para}
    \centering

    \begin{tabular}{c|c|c|c}
    \hline

    \hline
    From-node \# & To-node \# & $R$ (p.u.) & $X$ (p.u.)\\
    \hline
        $18$& $135$ & $1.2030$ & $1.1034$\\
    \hline
        $13$& $152$ & $1.0300$ & $0.7400$\\
    \hline
       $151$& $300$ & $1.4512$&$1.3083$\\
    \hline
        $54$& $94$ & $1.5042$&$1.3554$\\
    \hline
        $97$& $197$ & $1.4680$&$1.1550$\\
    \hline

    \hline
    \end{tabular}
\end{table}
\begin{table}
    \caption{Control Parameters, Pre-event Measurements and Post-event Setpoints of the IEEE 123-node Feeder}
    \label{tab:123_ctrl_para}
    \centering

    \begin{tabular}{c|c|c|c|c|c}
    \hline

    \hline
    &MG1& MG2&MG3&MG4&MG5\\
    \hline
    $M_{\text{a}k}$&$1.2$ &$1$&$0.8$&$1$&$1.2$\\
    \hline
    $D_{\text{a}k}$&$1.2$ &$1.2$&$1.2$&$1.2$&$1.2$\\
    \hline
    $M_{\text{v}k}$&$12$&$10$&$16$&$10$&$12$\\
    \hline
    $D_{\text{v}k}$&$0.2$&$0.2$&$0.2$&$0.2$&$0.2$\\
    \hline
    Pre-event $\delta_k$ (rad.)&$0$& $-0.8472$ & $2.3062$ & $0.5936$&$0.7732$\\
    \hline
    $\delta_k^*$ (rad.)&$0$& $-1.0472$ & $2.3562$ & $0.5236$&N/A\\
    \hline
    $E_k^*$ (p.u.)& $1.0$&$1.0$&$1.0$&$1.0$&N/A\\
    \hline

    \hline
    \end{tabular}
\end{table}

\subsubsection{Online Application of Estimated security region} 
\label{ssub:online_application_of_estimated_security_region}
\begin{figure}
        \centering
        \subfloat[]{\includegraphics[width=1.35in]{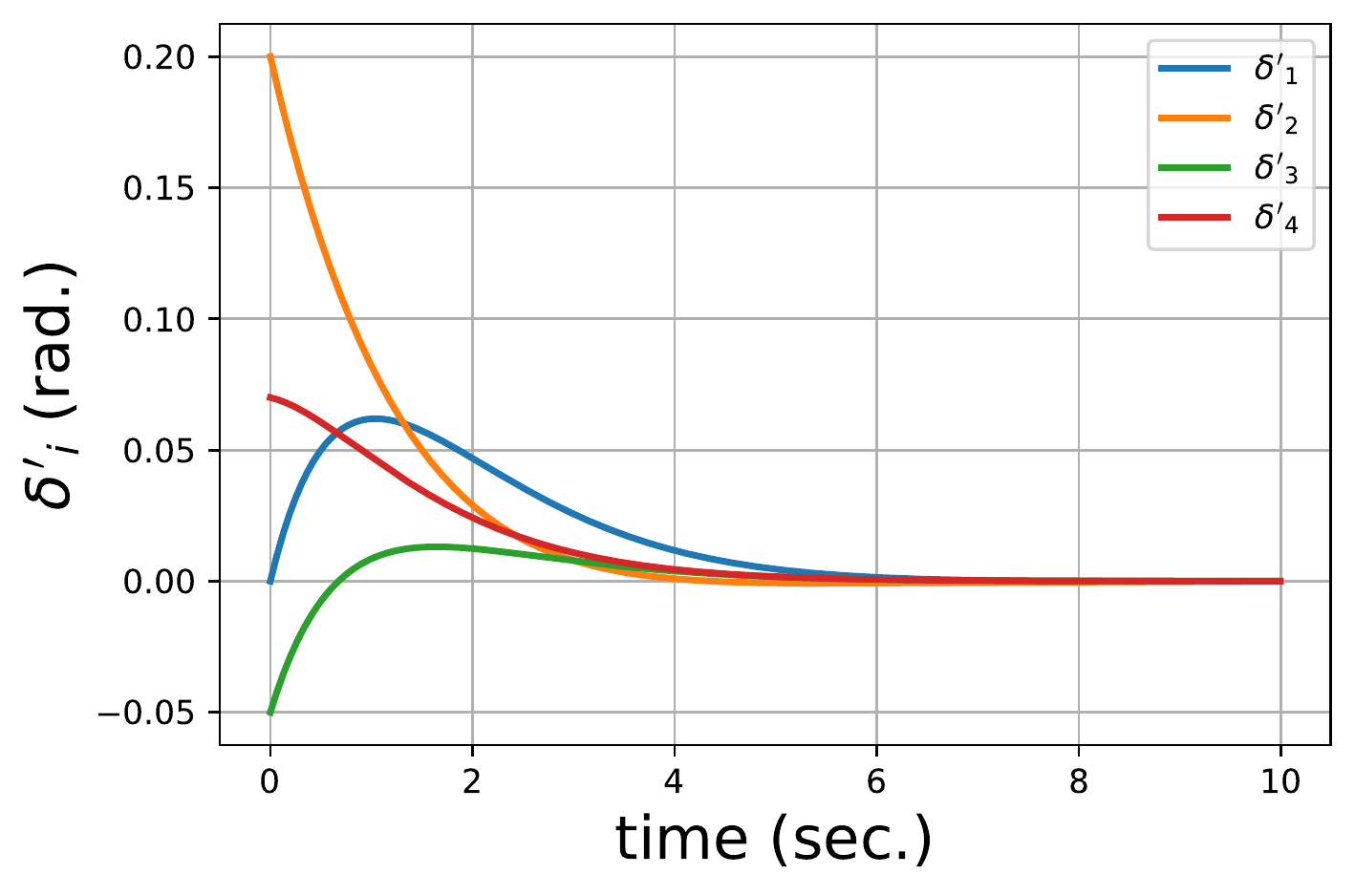}}
        \hfil
        \subfloat[]{\includegraphics[width=1.35in]{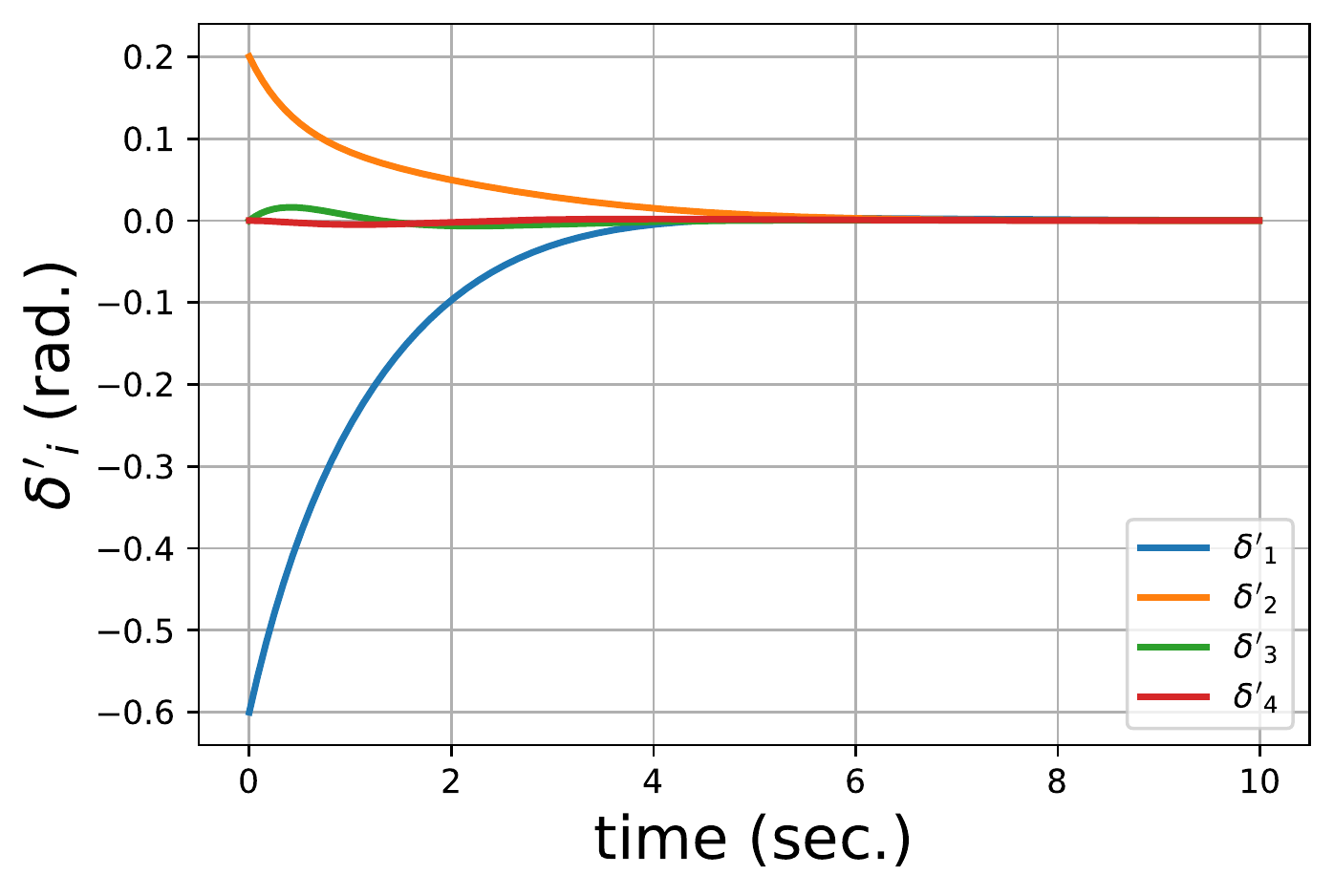}}
        \hfil
        \caption{Time-domain simulation of interface variables in the 123-node feeder: (a) with MG 5 islanded; (b) with $\mathbf{x}(0) = [-0.6, 0.2, 0, 0]^{\top}$ rad.}
        \label{fig:FiveMG_Time_domain}
    \end{figure}

Suppose that at time $t=0$, MG $5$ enters an islanded mode and the DSO would like to know if the remaining $4$ networked MGs can be stabilized at a pre-dispatched operating point. During offline planning, Algorithm \ref{alg:security_region_estimation} computes a Lyapunov function $V_{\boldsymbol{\theta}^*}$ and a security region $\mathcal{S}_{0.69}$ for the contingency. $\mathcal{S}_{0.69}$ can be leveraged during real-time operation, in order to determine if the remaining MGs can tolerate the disturbance due to islanding of MG 5. The initial condition $\mathbf{x}(0)$ can be obtained by collecting pre-event measurements at the MG interfaces. In this case study, $V_{\boldsymbol{\theta}^*}(\mathbf{x}(0))=0.12<0.69$, suggesting that $\mathbf{x}(0) \in \mathcal{S}_{0.69}$. Thus, without any simulation, the DSO can almost instantaneously conclude that all interface variables tend to their pre-dispatched values. Such a conclusion is confirmed by the time-domain simulation in Figure \ref{fig:FiveMG_Time_domain}-(a).
\subsubsection{Learned Lyapunov Function and Estimated Security Region} 
\label{ssub:learned_lyapunov_function_and_estimated_security_region}
\begin{figure}
        \centering
        \subfloat[]{\includegraphics[width=1.4in]{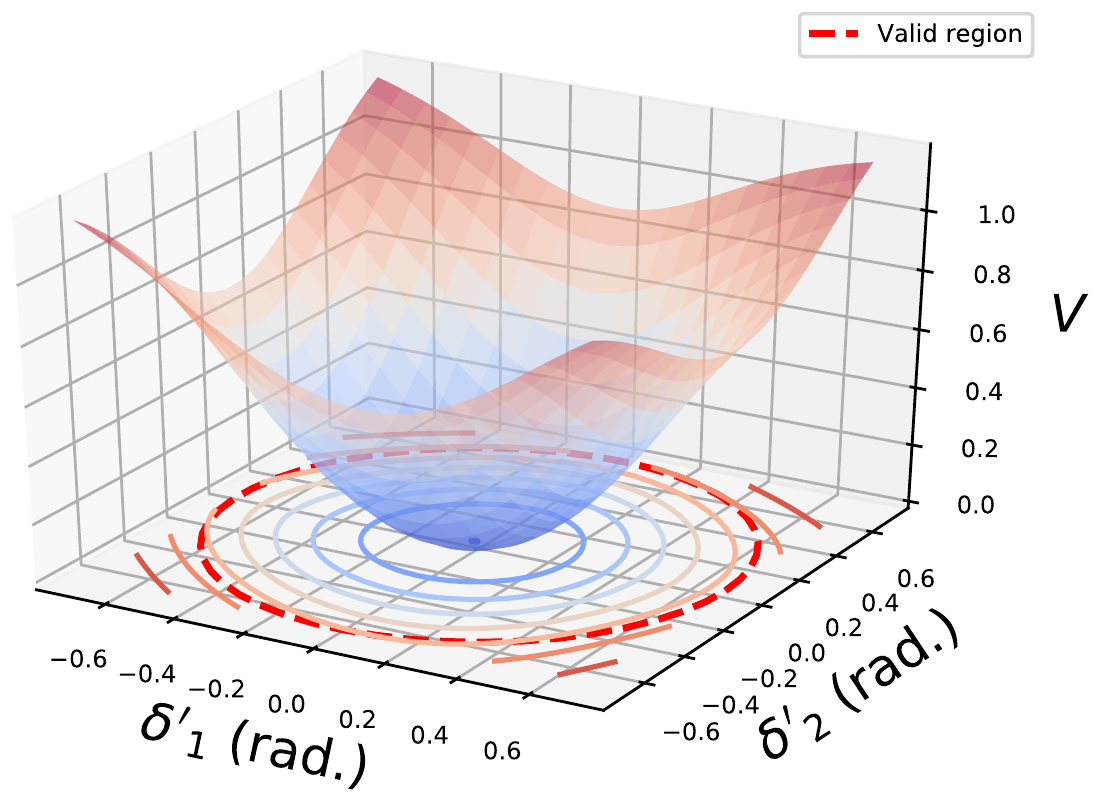}}
        \hfil
        \subfloat[]{\includegraphics[width=1.4in]{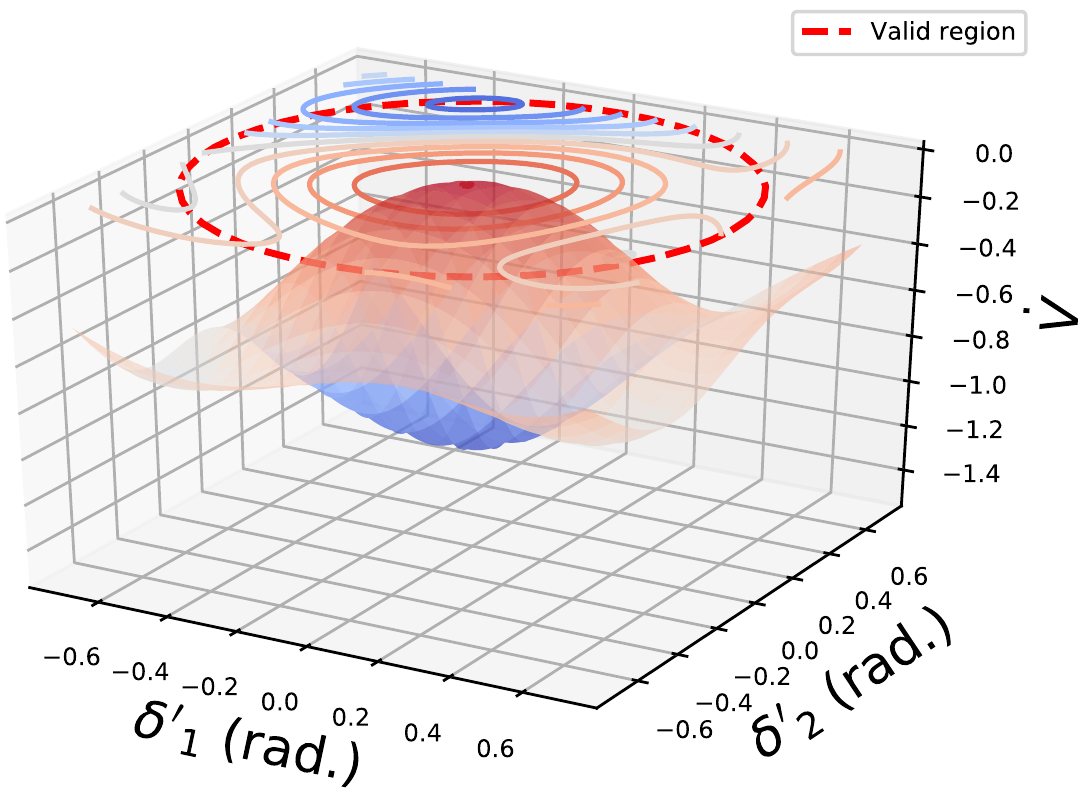}}
        \hfil
        \caption{(a) $V_{\boldsymbol{\theta}^*}$ and (b) $\dot{V}_{\boldsymbol{\theta}^*}$ in the 123-node feeder.}
        \label{fig:FiveMG_LF}
    \end{figure}

\begin{figure}
        \centering
        \subfloat[]{\includegraphics[width=1.35in]{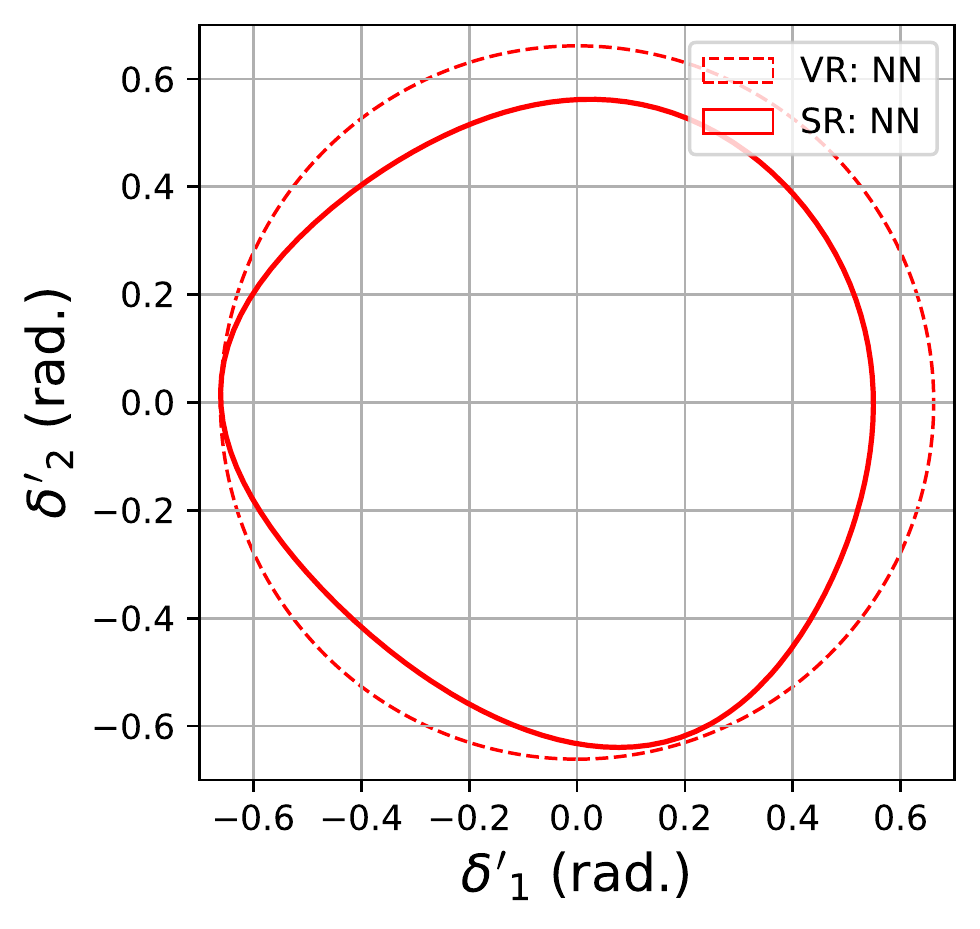}}
        \hfil
        \subfloat[]{\includegraphics[width=1.35in]{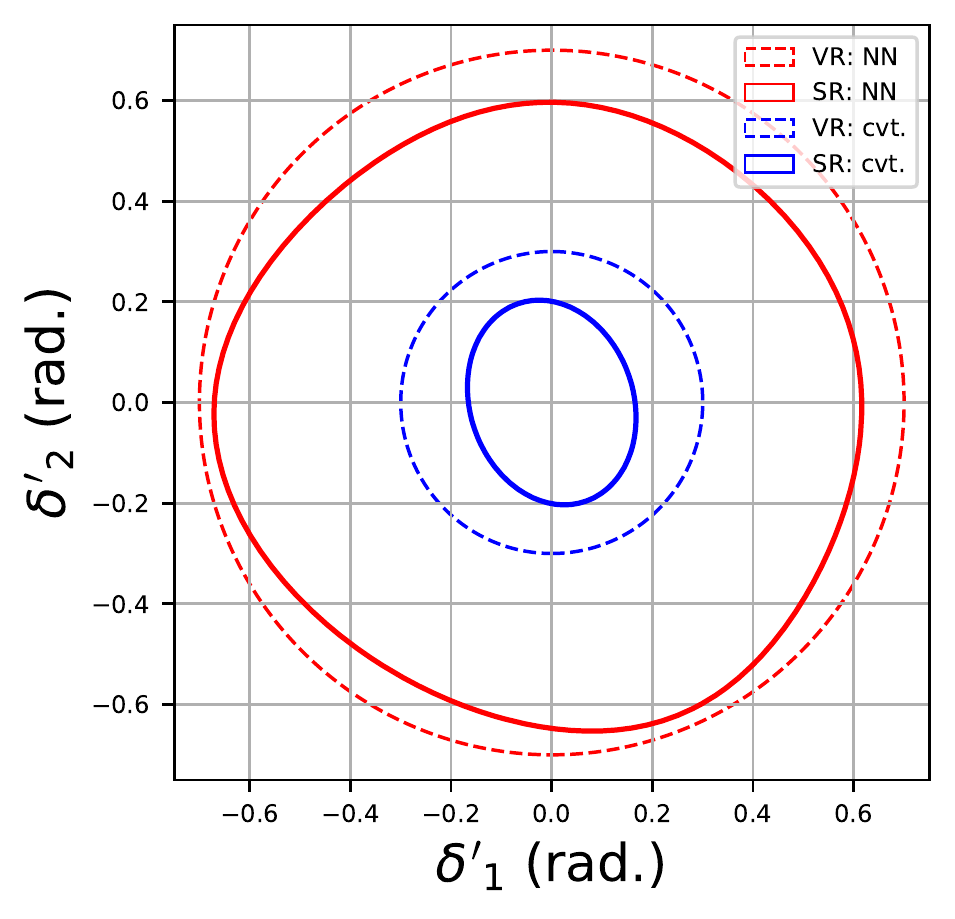}}
        \hfil
        \caption{(a) security region (SR) and valid region (VR) around the touching point with $\delta'_3=0.06$ and $\omega'_3=0.22$; (b) comparison between the proposed (NN) and conventional (cvt.) methods with $\delta'_3=\omega'_4=0$.}
        \label{fig:FiveMG_Compare}
    \end{figure}

It takes $2901.69$ seconds to learn the Lyapunov function $V_{\boldsymbol{\theta}^*}$. Figure \ref{fig:FiveMG_LF} visualizes $V_{\boldsymbol{\theta}^*}$ and $\dot{V}_{\boldsymbol{\theta}^*}$. With $V_{\boldsymbol{\theta}^*}$, \texttt{SREst} computes an security region which is visualized in Figure \ref{fig:FiveMG_Compare} and it suggests that the solution to \eqref{eq:d_star_original} is $[-0.66, 0.03, 0.06, 0.22]^{\top}$. Figure \ref{fig:FiveMG_Compare}-(a) visualizes the region $\mathcal{S}_{0.69}$ in the $\delta'_1$-$\delta_2'$ plane with $\delta'_3 = 0.06$ and $\delta'_4 = 0.22$. It is observed that the touching point of the boundaries of $\mathcal{S}_{0.69}$ and $\mathcal{B}_{0.7}$ is $(0.66, 0.03)$.

%

\subsubsection{Comparison} 
\label{ssub:comparison_study}
The comparison between the security region estimated based on the proposed and conventional methods is shown in Figure \ref{fig:FiveMG_Compare}-(b). Denoted by $\mathcal{S}'''$ is the security region estimated based on the conventional approach. Suppose that pre-event operating condition $\mathbf{x}(0)$ is $[-0.6, 0.2, 0,0]^{\top}$. Such a condition is outside $\mathcal{S}'''$ but inside $\mathcal{S}_{0.69}$. Therefore, $\mathcal{S}_{0.69}$ can conclude that \emph{the system trajectory will converge to the equilibrium whereas $\mathcal{S}'''$ cannot}. The time-domain simulation shown in Figure \ref{fig:FiveMG_Time_domain}-(b) confirms the convergence of the states given the pre-event condition.

\section{Conclusion} 
\label{sec:conclusion}
In this paper, we propose a TSA tool for networked microgrids based on Neural Lyapunov Methods. Assessing transient stability is formulated as a problem of estimating the security region of networked microgrids. We use neural networks to learn a Lyapunov function in the state space. The optimal security region is estimated based on the function learned, and it can be used for both offline design and online operation. The effectiveness of the proposed TSA tool is tested and validated in 3 scenarios: 1) a grid-connected microgrid, 2) a three networked microgrids with heterogeneous dynamics, and 3) the IEEE 123-node test feeder. In the proposed TSA tool, the SMT solver is used to augment the training set, which is computationally expensive. Future work will develop more efficient algorithms to speed up the procedure of learning a Lyapunov function in larger systems.

\bibliographystyle{IEEEtran}
\bibliography{ref_short}







\end{document}